\documentclass[usenatbib]{aa}
\usepackage{times}
\usepackage{graphicx}
\usepackage{natbib}
\usepackage{subfigure}
\usepackage{dcolumn}
\usepackage{longtable}
\usepackage{lscape}
\usepackage{ulem} 
\usepackage[usenames,dvipsnames]{xcolor}
\usepackage{xspace}
\usepackage{aas_macros}
\usepackage{xargs}
\usepackage{ulem}
\usepackage{url}

\newcolumntype{d}{D{.}{.}{-1}}

\title{The physical properties of Spitzer/IRS galaxies derived from their UV to 22 $\mu$m spectral energy distribution}
\author{Marina~Vika\inst{\ref{inst1}}
\and Laure~Ciesla\inst{\ref{inst2},\ref{inst3}}
\and Vassilis~Charmandaris\inst{\ref{inst1},\ref{inst3}}
\and Emmanuel M.~Xilouris\inst{\ref{inst1}}
\and Vianney Lebouteiller\inst{\ref{inst2}}
}

\institute{Institute for Astronomy, Astrophysics, Space Applications \& Remote Sensing, National Observatory of Athens, Penteli, 15236, Athens, Greece\label{inst1} \email{mvika@astro.noa.gr}
\and Laboratoire AIM-Paris-Saclay, CEA/DSM/Irfu - CNRS - Université Paris Diderot, CEA-Saclay, F-91191 Gif-sur-Yvette, France  \label{inst2}
\and   University of Crete, Department of Physics, Heraklion 71003, Greece\label{inst3}
}

\begin{document}

\date{Received .../ 
Accepted ...}

\label{firstpage}

\abstract{
We provide the basic integrated physical properties of all the galaxies contained in the full Cornell Atlas of Spitzer/IRS Sources (CASSIS) with available broad-band photometry from UV to 22 $\mu$m. 
We have collected broad-band photometric measurements in 14 wavelengths from available public surveys in order to study the spectral energy distribution (SED) of each galaxy in CASSIS, thus constructing a final sample of 1,146 galaxies in the redshift range $0 < \rm z < 2.5$.
The SEDs are modelled with the CIGALE code which relies on the energy balance between the absorbed stellar and the dust emission while taking into account the possible contribution due to the presence of an active galactic nucleus (AGN).
We split the galaxies in three groups, a low-redshift (z$<0.1$), a mid-redshift ($0.1 \leq \rm z<0.5$) and a high-redshift (z$ \geq 0.5$) sub-sample and find that the vast majority of the Spitzer/IRS galaxies are star-forming and lie on or above the star-forming main sequence of the corresponding redshift.
Moreover, the emission of Spitzer/IRS galaxies with z$<0.1$ is mostly dominated by star-formation, 
galaxies in the mid-redshift bin are a mixture of star forming and AGN galaxies, while half of the galaxies with z$ \geq 0.5$ show moderate or high AGN activity. 
Additionally, using rest-frame $NUV-r$ colour, S\'ersic indices, optical [O$_{\rm III}$] and [N$_{\rm II}$] emission lines we explore the nature of these galaxies by investigating further their structure as well as their star-formation and AGN activity.
Using a colour magnitude diagram we confirm that 97\% of the galaxies with redshift smaller than 0.5 have experienced a recent star-formation episode.
For a sub-sample of galaxies with available structural information and redshift smaller than 0.3 we find that early-type galaxies are placed below the main sequence,
while late-type galaxies are found on the main-sequence as expected.
Finally, for all the galaxies with redshift smaller than 0.5 and available optical spectral line measurements we compare the ability of CIGALE to detect the presence of an AGN in contrast to the optical spectra classification.
We find that galaxies with high AGN luminosity, as calculated by CIGALE, are most likely to be classified as composite or AGNs by optical spectral lines.
}


\titlerunning{CASSIS SEDs}
\authorrunning{Vika et al.}
\maketitle

\section{Introduction}
\label{sec:intro}

Understanding how galaxies obtain their baryonic matter over cosmic time is an open question in extragalactic astronomy. 
Multiple physical processes can influence the star-formation history (SFH) of a galaxy, including external processes such as minor mergers, gas accretion, dynamical heating of stellar populations, as well internal processes, including massive wind outflows or feedback due to active galactic nuclei (AGN). 
For constraining the star-formation history and explaining the evolution of their baryonic content we need accurate measurements of physical parameters in particular stellar masses ($M_{\star}$), 
star-formation rates (SFR), dust content together with measurements of the possible AGN contribution to the total galaxy luminosity at different epochs of the galaxy evolution.

The spectral energy distribution (SED) of a galaxy, typically estimated by collecting broad band photometry across the all possible wavelengths, 
is a valuable source from which one can extract key physical properties of the unresolved stellar population (see \citet{tex:WG11} for a review). 
A SED comprises the emission from the stars as well as the interstellar gas and dust, with stars emitting mainly in the UV-optical wavelengths while dust absorbing part of the stellar light and re-emiting it at infrared (IR) and the submillimeter  wavelengths. 
Since the UV-optical absorbed energy is re-emitted up to submillimeter wavelengths, the intrinsic stellar emission can be constrained by gathering observations from UV to far-IR after applying energy balance arguments (i.e. \citealt{tex:dC08,tex:NB09}). 
When a galaxy hosts an accreting supermassive black hole,
 the emission from the central active galactic nucleus (AGN) may also contribute to the global optical and IR power output of the galaxy and should be taken into account in order to properly estimate both the stellar mass and the star-formation rate (\citealt{tex:CC15}). 

In addition to the SED analysis, moderate or high resolution spectroscopy can provide even more information on galaxy physical properties. 
One can use detailed theoretical predictions regarding the strength of fine-structure lines or molecular spectral features to infer the details of excitation mechanisms, chemical composition, strength of radiation field, and amount of dust extinction. 
In particular mid-and far-IR spectroscopy, even though challenging to obtain, is an extremely powerful probe of the nuclear activity, 
since it's less affected by obscuration and samples a wealth of ionic and rotational/vibrational features (\citealt{tex:CL01, tex:D03}).

In the present work, we use as a basis for our study the Cornell AtlaS of Spitzer/IRS Sources (CASSIS\footnote{http://cassis.astro.cornell.edu/atlas/}; \citealt{tex:LB11b,tex:LB15}), 
which contains all pointed observations obtained by the Infrared Spectrograph (IRS; \citealt{tex:HR04b}) on board the Spitzer Space Telescope (\citealt{tex:WR04}). 
We select all the galaxies for which in addition to the 5-37$\mu$m mid-IR spectrum, ancillary broad band photometry from UV to mid-IR (22$\mu$m) is publicly available.
The main goal of this paper is to present a panchromatic atlas of the broadband SEDs of these galaxies and to provide their global properties such as stellar mass, star-formation rate, AGN luminosity,
as well as the contribution of an AGN to the total IR luminosity (${\rm frac}_{\rm AGN}$), as derived via a global SED modelling.

It is envisioned that the availability of Spitzer/IRS mid-IR spectroscopy for this infrared selected sample will enable us to better constrain their energy production mechanism in the often dust enshrouded nuclear regions. 
The Spitzer/IRS observations provide the 5-37$\mu$m galaxy emission with the current highest spatial resolution that allows for new mid-IR diagnostics to be developed, something that was not possible with previous shallow mid-IR spectra. 
So far, Spitzer/IRS spectra have enabled the detailed study of the various AGN types based on silicate (9.7$\mu$m, 18$\mu$m) and polycyclic aromatic hydrocarbons (PAHs, 6.2$\mu$m) emissions (e.g. \citealt{tex:WH10,tex:HH11,tex:SY12,tex:SA14}).
It has been shown that silicate features can vary with the AGN type (\citealt{tex:SM07,tex:HW07,tex:HH15}) while in case of early-type galaxies they provide a strong diagnostic tool for the population content of the galaxies (\citealt{tex:BP06b}).
For the case of the luminous and ultra-luminous IR galaxies (LIRGs/ULIRGs), the PAHs and silicate features allowed the further classification of the various subtypes of infrared galaxies  (\citealt{tex:SA13} and references therein).
Furthermore, Spitzer/IRS spectra can provide an additional method of separating the nuclear emission from AGNs and star-formation when both are present in a galaxy (e.g. \citealt{tex:PA11}).
Thus, this work aims to produce a complete catalogue of stellar masses, SFRs, 
stellar ages, together with measurements of the dust content and AGN luminosity of all the Spitzer/IRS observed galaxies that will allow for future studies to compare the mid-IR spectral properties with the globally derived ones from the galaxy SED.

In Section 2 we describe the sample construction. 
In Section 3 we present the methodology used in order to measure the global physical properties. 
In Section 4 we explore the nature of the galaxies found in the CASSIS sample, while our conclusions are presented in Section 5.  
Throughout this paper we use $H_{0} = 70$, $\Omega_{\rm M}=0.3$ and $\Omega_{\Lambda}=0.7$. All magnitudes are given in AB system.

\section{Sample Construction}
\label{sec:data}

\begin{table}
\caption{Photometric bands and number of sources used in our study.}
  \smallskip
  \centering
  \begin{tabular}{cccc}
  \hline
  \noalign{\smallskip}
Survey & Band  & Wavelength ($\mu$m) &  \# of sources     \\  
\noalign{\smallskip}
\hline
 \noalign{\smallskip}
\hline
 \noalign{\smallskip}
GALEX               &  $FUV$ &  0.152    &   892  \\
GALEX               &  $NUV$ &  0.227    &  1146  \\
SDSS                &  u   &  0.359    &  1087  \\
SDSS                &  g   &  0.463    &  1089  \\
SDSS                &  r   &  0.614    &  1088  \\
SDSS                &  i   &  0.747    &  1088  \\
SDSS                &  z   &  0.893    &  1089  \\
2MASS/UKIRT 		&  J   &  1.24/1.25  &   926  \\
2MASS/UKIRT 		&  H   &  1.66/1.63  &   926  \\
2MASS/UKIRT         &  K   &  2.16/2.2   &   934  \\
WISE                &  W1  &  3.37     &  1146  \\
WISE                &  W2  &  4.61     &  1146  \\
WISE                &  W3  &  12.08    &  1141  \\
WISE                &  W4  &  22.19    &  1120  \\
\noalign{\smallskip}
\hline
\end{tabular}
\tablefoot{Column 1: Survey name, Column 2: bandpass name, Column 3: effective wavelength of each passband, Column 4: number of galaxies with flux measurements in each band. 
See Section~\ref{sec:data} for the requirement of keeping a galaxy in the final sample.}
\label{table:table1} 
\end{table}

\begin{figure}[!htb]
\centering
\includegraphics[width=0.5\textwidth]{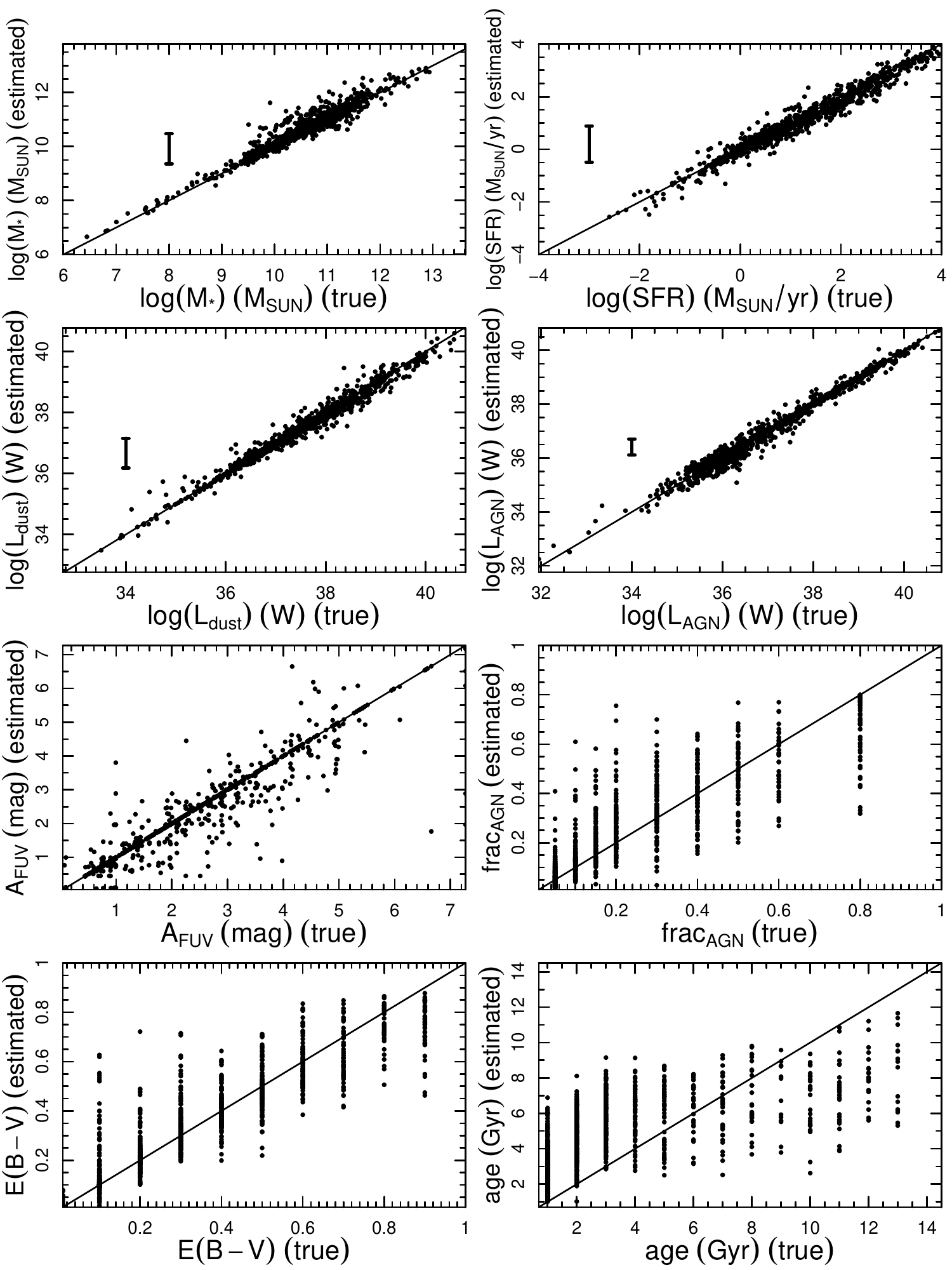}
\caption{Comparison between the true values of the parameters and the mock catalogs parameter values estimated from the probability distribution function of each parameter. The solid black line is the 1:1 relation.}
\label{fig:mock}
\end{figure}

\begin{table*}
	\caption{The CIGALE input parameters and their ranges whenever appropriate.}
	\smallskip
    \centering
	\begin{tabular}{l c }
	\hline
    \noalign{\smallskip}
	Parameter & Value \\ 
	\noalign{\smallskip}
\hline
 \noalign{\smallskip}
\hline
 \noalign{\smallskip}
        SFH         			&  Delayed 		        \\  
	age (Gyr) 			& 0.1, 0.5, 1, 2, 3, 4, 5, 	\\
	                			& 6, 7, 8, 9, 10, 11, 12, 13	\\
	e-folding time in Gyr ($\tau_{main}$) & 0.5, 1, 3, 5, 6, 10, 20\\
	Possible quenching or burst ($r_{\rm SFR}$) & 0., 0.1, 0.2, 0.3, 0.4, 0.5, 0.6, \\
    					& 0.7, 0.8, 0.9, 1., 2., 4., 8., 16.\\
	Duration of the episode in Gyr & 0.1, 0.4 \\				
\hline
	Stellar Population Models & \citet{tex:BC03}	\\
	IMF & Salpeter\\
	Metallicity & 0.02\\
\hline
	Dust attenuation $E(B-V)$	&  0.01, 0.05, 0.1, 0.15, 0.2, 0.25, 	\\
	        	&  0.3, 0.35, 0.4, 0.45, 0.5, 0.6, 0.7, 0.9	\\
\hline		
	Dust template & \citet{tex:DH14b}\\
\hline
	IR power-law slope ($\alpha$\tablefootmark{b})	& 	1.5, 2.5  \\  
\hline
     AGN template & \citet{tex:FF06} \\
\hline
	Ratio between outer and inner radius of the torus ($r_{ratio}$)	& 	60  \\
	Optical depth at 9.7 $\mu$m ($\tau$)	& 	1, 6  \\
	Parameter linked to the radial dust distribution in the torus ($\beta$)	& 	-0.5  \\
	Parameter linked to the angular dust distribution in the torus ($\gamma$)	& 	0  \\
	Angular opening angle of the torus ($\theta$)	& 	100  \\
	Angle with line of sight ($\psi$)	& 	0, 90  \\
	${\rm frac}_{\rm AGN}$	& 	0.0, 0.05, 0.1, 0.15, 0.2,  \\
	            	& 	0.3, 0.4, 0.5, 0.6, 0.8 \\
 \hline
       Number of models per redshift bin (Delta(z) = 0.01): & 2808000\\
	\noalign{\smallskip}
\hline
\end{tabular}
    \label{fitparam}
\end{table*}

The CASSIS spectroscopic sample consists of 15,729 spectral observations \citep{tex:LB15}. 
The first step was to select all the sources with an extragalactic identification as defined by CASSIS\footnote{Keyword Category in http://isc.astro.cornell.edu/Smart/ProgramIDs}, including all the sources with unknown identification. 
This selection limits our sample to 7,601 spectral observations. 
The sample of 7,601 pointings was cross-matched with various catalogues that contain broad band photometry from $FUV$ to mid-IR. 
More specifically, we searched for photometric counterparts in GALEX (all-sky survey - AIS; median - MIS; nearby galaxy survey - NGS, \citet{tex:GB07,tex:BE11}), SDSS (DR10, tex:AA14), 2MASS (XSC; PSC, \citet{tex:CS03}), UKIRT (DR2, \citet{tex:WC07}) and WISE (All sky survey; WISE-SDSS, \citet{tex:JC00,tex:LH15}) within a radius of 3 arcsec\footnote{The acronyms in the brackets indicates the data release used to access the broadband photometric fluxes.}.
Galaxies in the CASSIS dataset with multiple pointings were identified and selected once.
This brought the total number of selected galaxies down to 4,286.

We refer the reader to each data release paper for details of the method used to measure the flux. 
Briefly, aperture photometry was used in the following surveys: \textit{GALEX} (MIS, AIS), \textit{UKIRT} (DR2) and \textit{WISE} (All sky survey). 
However, SDSS (DR10) uses de Vaucouleurs or exponential profile, 2MASS (XSC) adopts a combination of adaptive aperture photometry or exponential profiles, and GALEX (NGS) measures asymptotic magnitudes. 
Finally, forced aperture photometry based on the SDSS profiles was used by the WISE-SDSS catalogue. 
In the cases where multiple measurements exist for one galaxy at a specific passband we choose the one that includes the total flux of the galaxy, especially for the extended sources.

Next, we searched for counterparts in the NASA Extragalactic Database (NED) for source identification and redshift measurements. 
The selection of the NED counterpart was done using the source type keyword of NED. 
If the closest counterpart is of type "G" or "QSO" we kept it, otherwise we searched for the closest counterpart with those keywords. 
If none was found within 3 arcsec, we kept the closest counterpart regardless of the keyword inside 3 arcsec. 
Almost all the redshift values used in this study are in agreement with the Infrared Database of Extragalactic Observables
from Spitzer (IDEOS) redshift catalogue of \citet{tex:HS16}.
There are seven cases
that our redshift measurements has more than 15\% difference compared to IDEOS catalogue, see notes of Table \ref{table:finalpar} for their IDs.
In all these cases \citet{tex:HS16} have selected an alternative way, e.g. derived from IRS spectra, to measure the redshift instead of using the NED provided values.
Note that in some cases \citet{tex:HS16} provides multiple redshift measurements for a single AORkey observation. 
The additional redshift measurements correspond to multiple galaxies observed in a single observation that happened to be along the long IRS slits during the same observation.

Additionally, NED provided us with galactic extinctions \citep{tex:SF11} and galaxy types.
The galactic extinctions are then used to correct the magnitudes from u-band to K-band.
The UV bands are provided already corrected in the original catalogues.

One requirement that we imposed for keeping a galaxy in the final sample was to have at least one UV flux ($FUV$ or $NUV$), 
optical or near-IR fluxes (minimum 3 bands in the 0.35-2.16 $\mu m$ wavelength range), 
WISE fluxes and a redshift measurement. 
The 1,252 galaxies that fulfilled the above condition were fitted with CIGALE.
By visually inspecting the modelled SEDs we found that this first requirement had to be expanded in order to select a well modelled sample thus we introduced two more requirements for keeping a galaxy in the final sample.
The second requirement restricts the sample further by excluding the cases that a galaxy has XSC photometry in 2MASS and lacks photometry from SDSS.
For these extended galaxies that lack optical photometry, the SED modelling is especially challenging since it assumes energy balance between bands with inhomogeneous photometry from one band to another and the measured fluxes can not be fitted well by the model.
Finally, the third requirement was set in order to reject galaxies with unusual SEDs that indicate either a problem with the cross-matching process or a problem with the measured photometry in some wavebands.
Considering these three requirements we concluded with a sample of 1,146 galaxies. 
In this sample there are 45 galaxies that their SDSS photometry has been flagged as saturated.
By visual inspection we found that 42 out of 45 galaxies do not show any sign of nuclear saturation. 
Instead there other neighbouring sources that are saturated with some spillover pixels passing through the disks of our galaxies. 
This does not affect the photometry on the image of the galaxy. 
All the 45 galaxies are flagged in the final provided Table.
In Table~\ref{table:table1} we tabulate the number of sources found in each survey for the final sample.
Clearly this sample is not complete, as it is biased towards galaxies sufficiently bright in the mid-IR to be observed with Spitzer/IRS.
However, it has the big advantage to be the largest sample currently possible that has both broadband measurements from UV to 22 $\mu$m and 5-35 $\mu$m spectra measurements.

A minimum of photometric uncertainties were added to all broadband measurements before the SED modelling with CIGALE.
The uncertainties are 0.3 mag for the $u$ and the W4 bands, 0.2 mag for the $z$, the three 2MASS bands and the W3 bands. 
For all remaining passbands a minimum uncertainty of 0.1 mag was adopted. 
In cases where a larger uncertainty was provided by a measurement we kept the larger value.
All the magnitudes were converted to the AB system and then to mJy for the SED fitting process.

\begin{figure*}
\centering
\includegraphics[height=6cm,width=6cm]{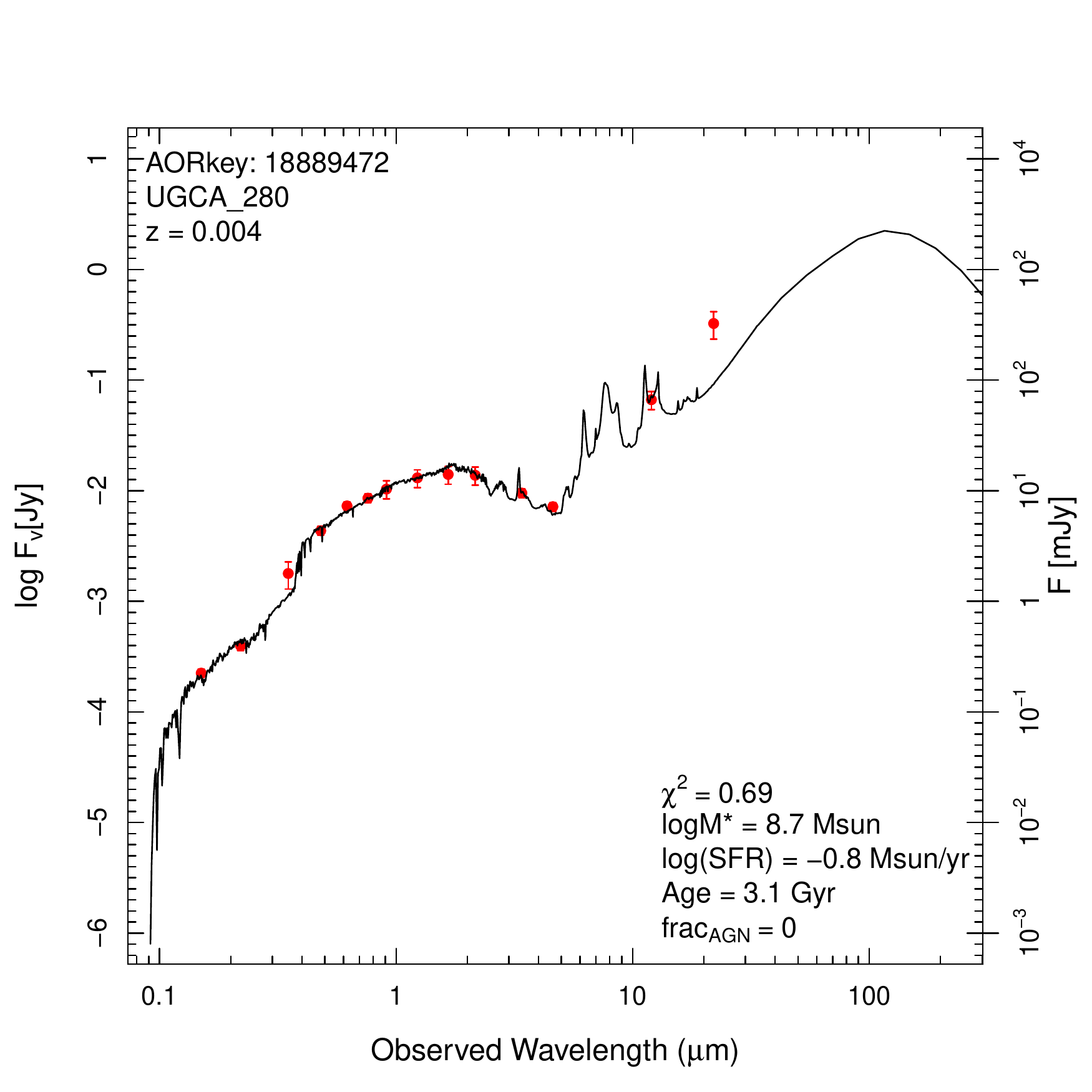}
\includegraphics[height=6cm,width=6cm]{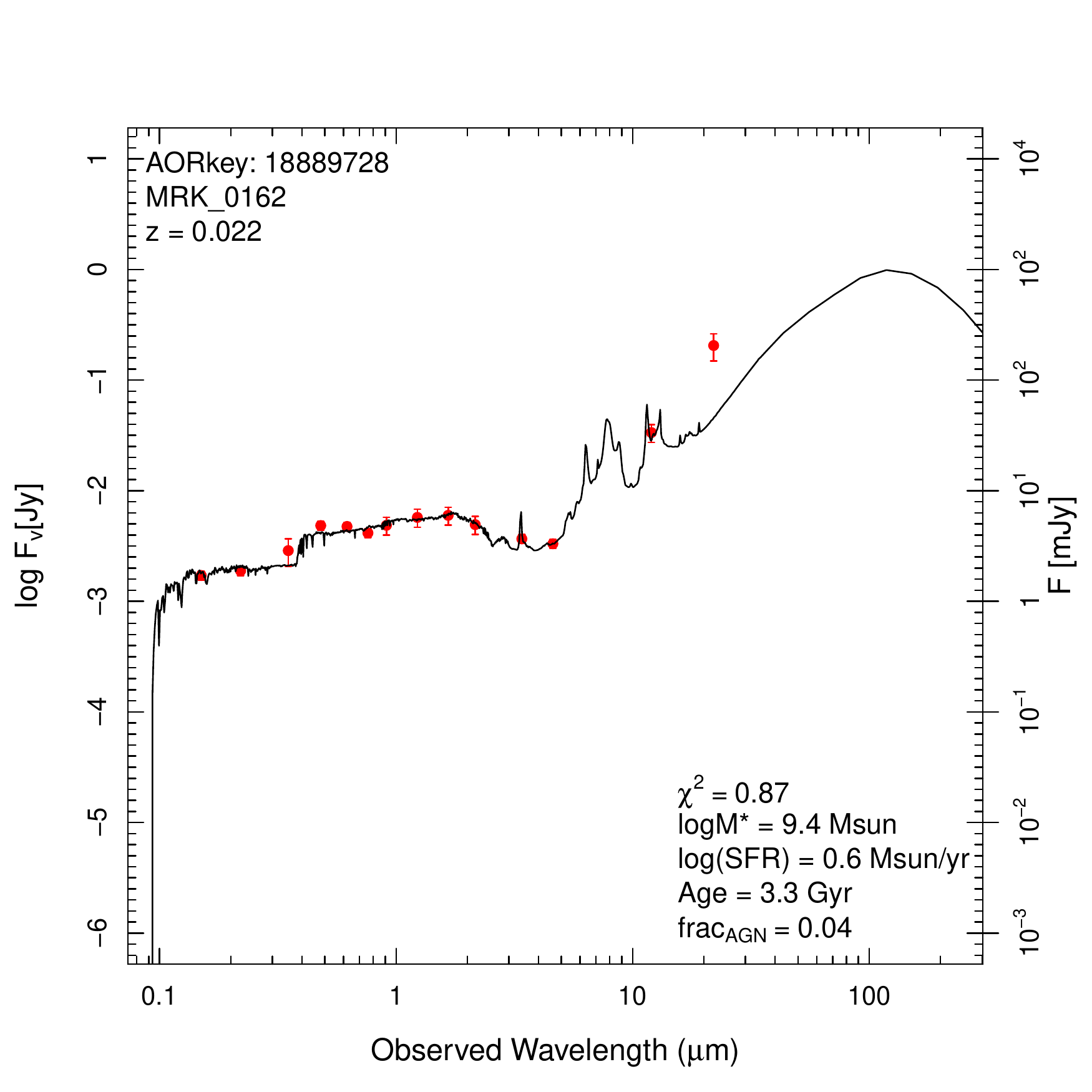}
\includegraphics[height=6cm,width=6cm]{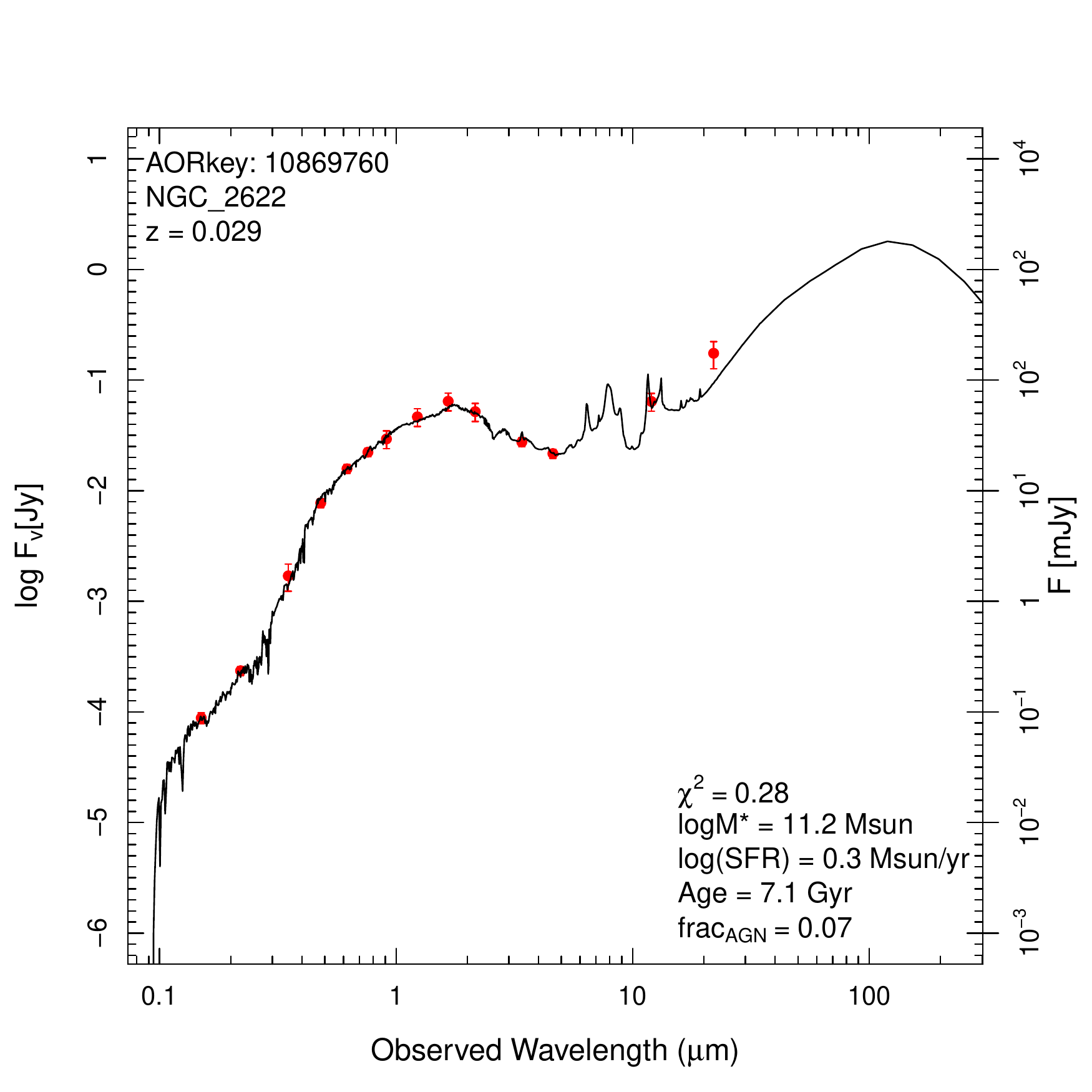}
\includegraphics[height=6cm,width=6cm]{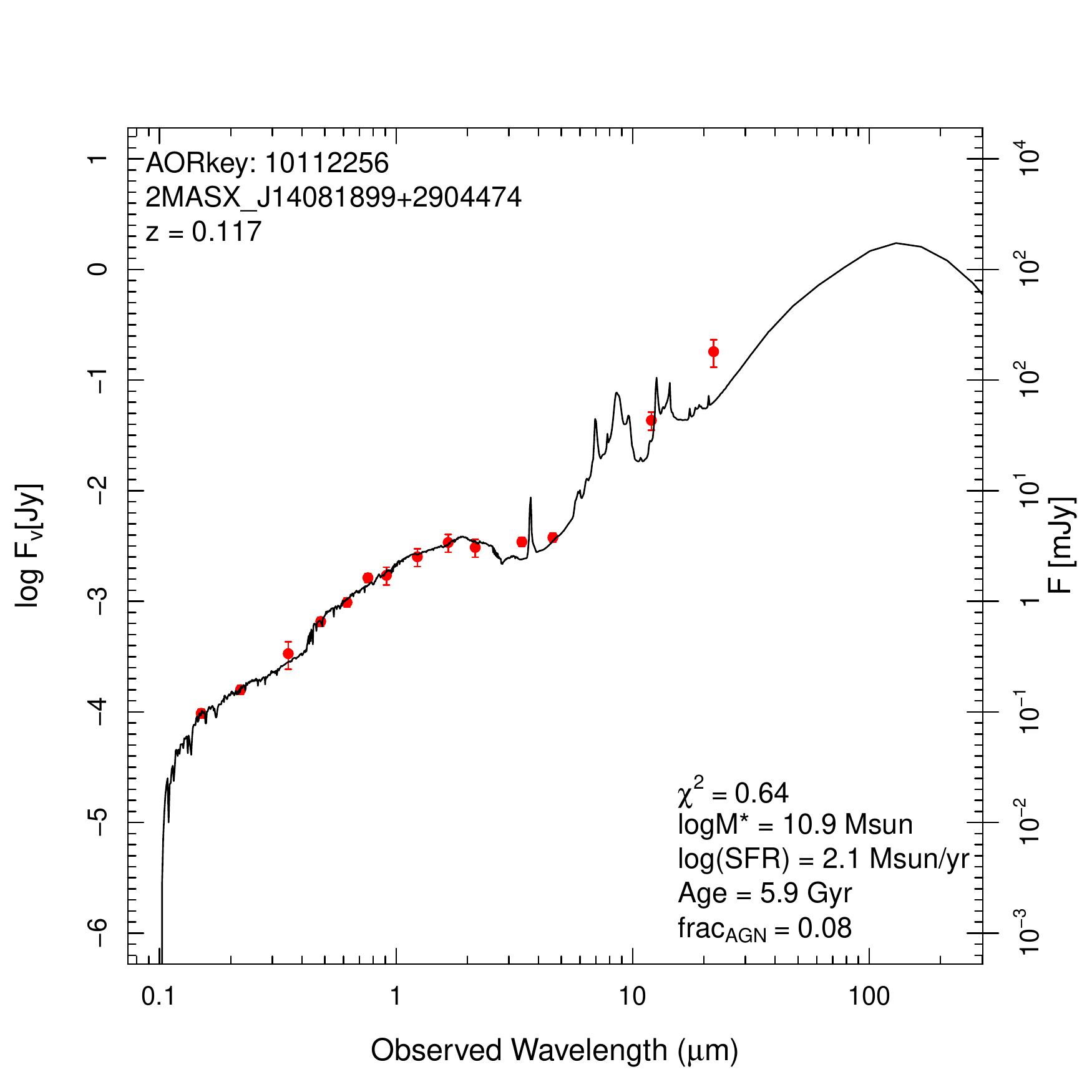}
\includegraphics[height=6cm,width=6cm]{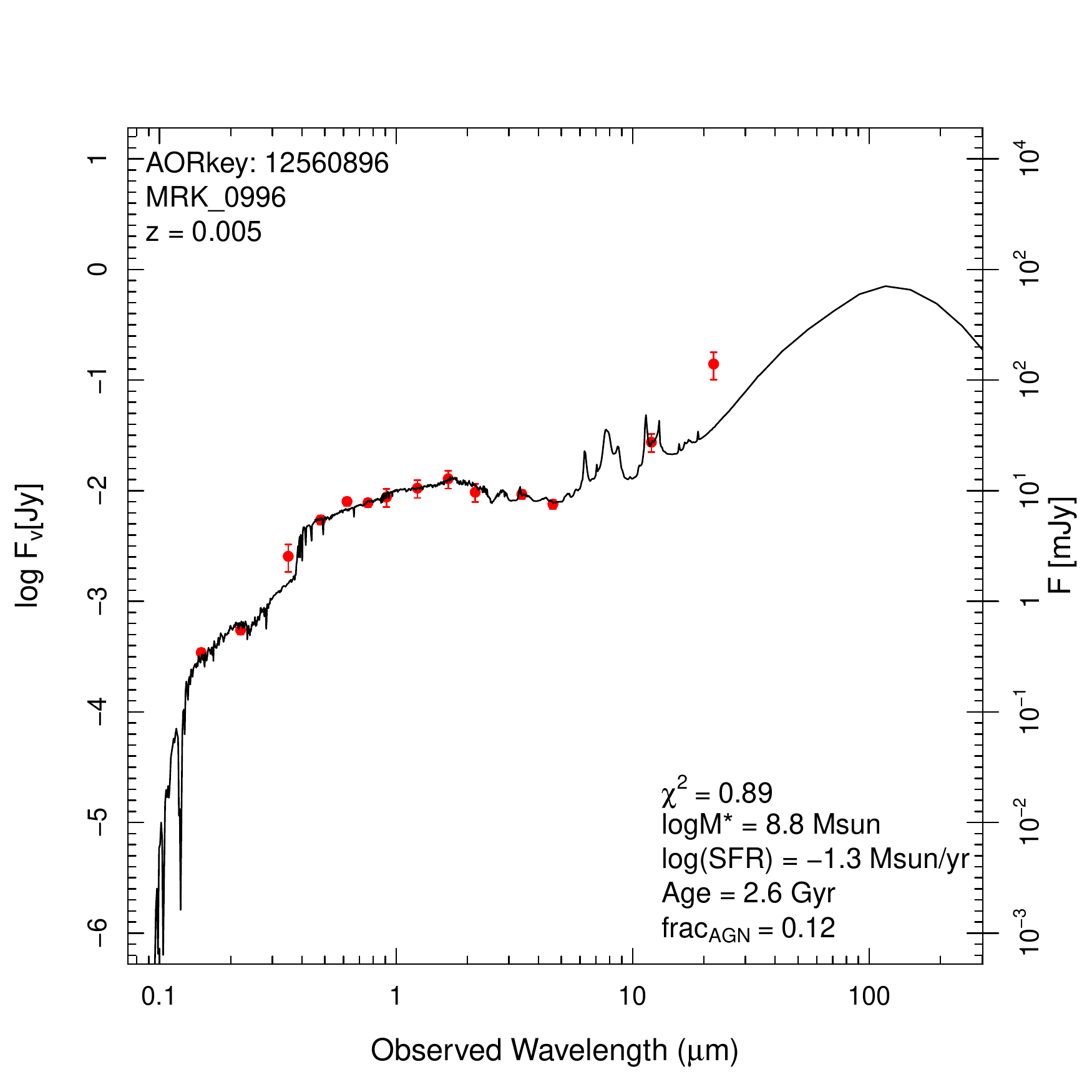}
\includegraphics[height=6cm,width=6cm]{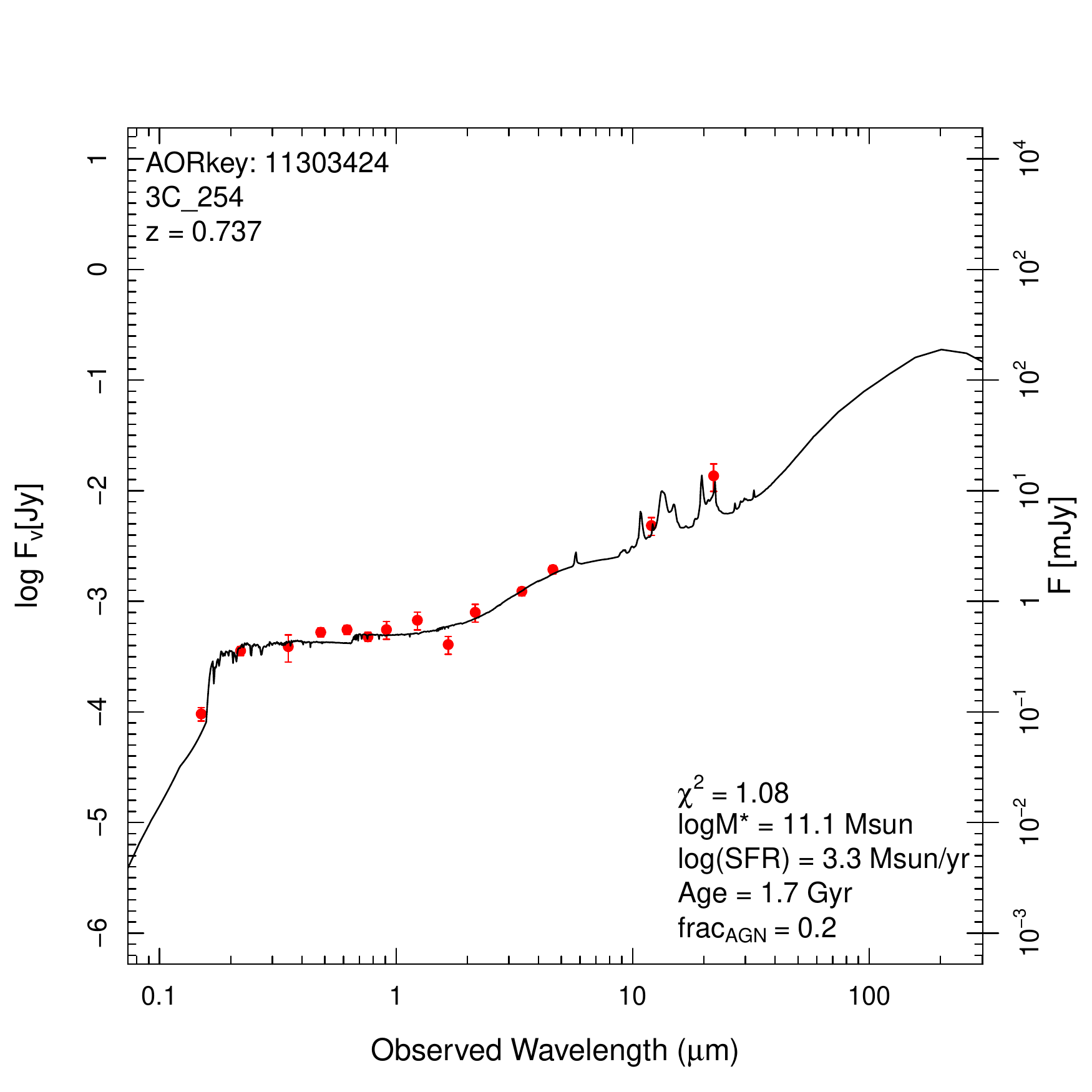}
\includegraphics[height=6cm,width=6cm]{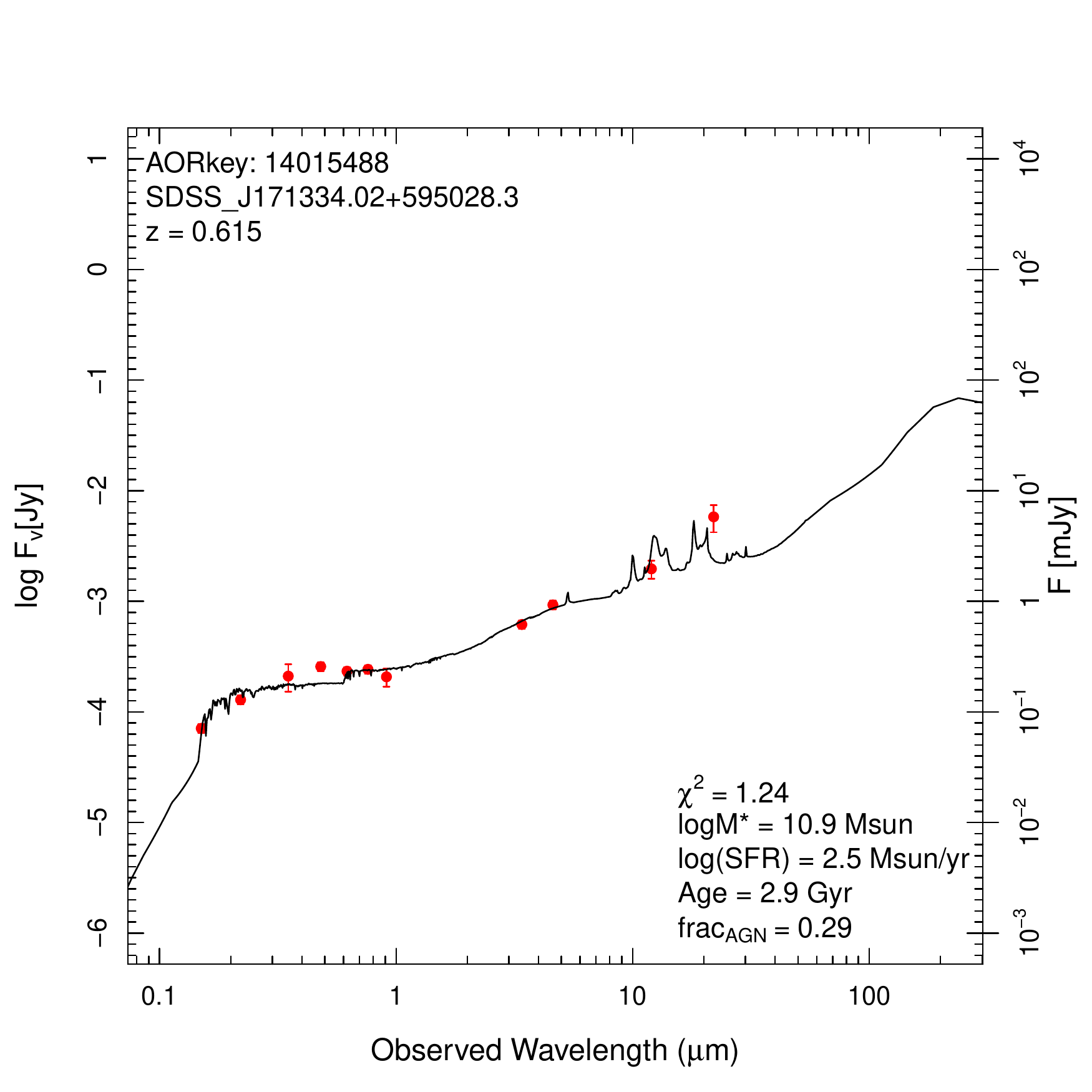}
\includegraphics[height=6cm,width=6cm]{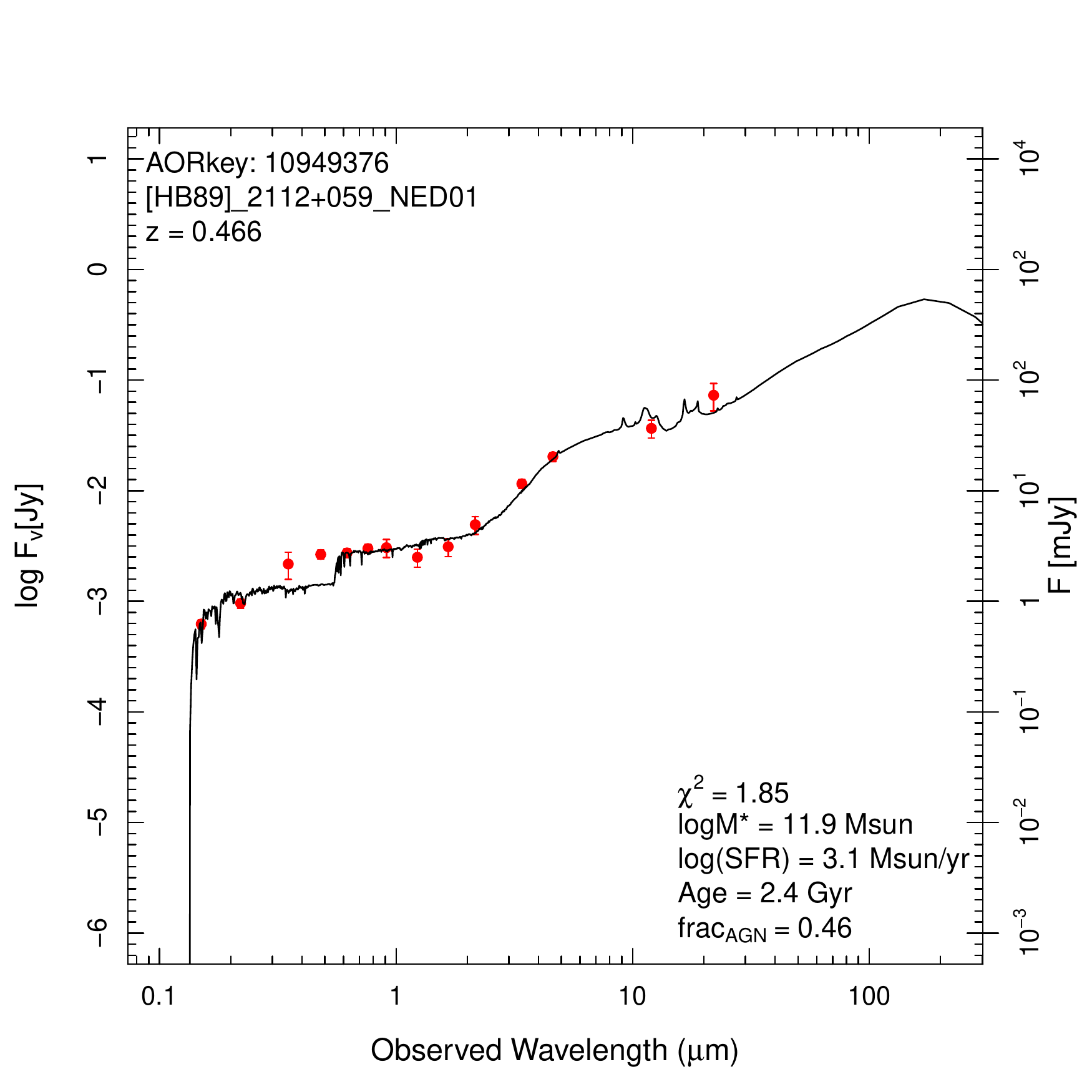}
\includegraphics[height=6cm,width=6cm]{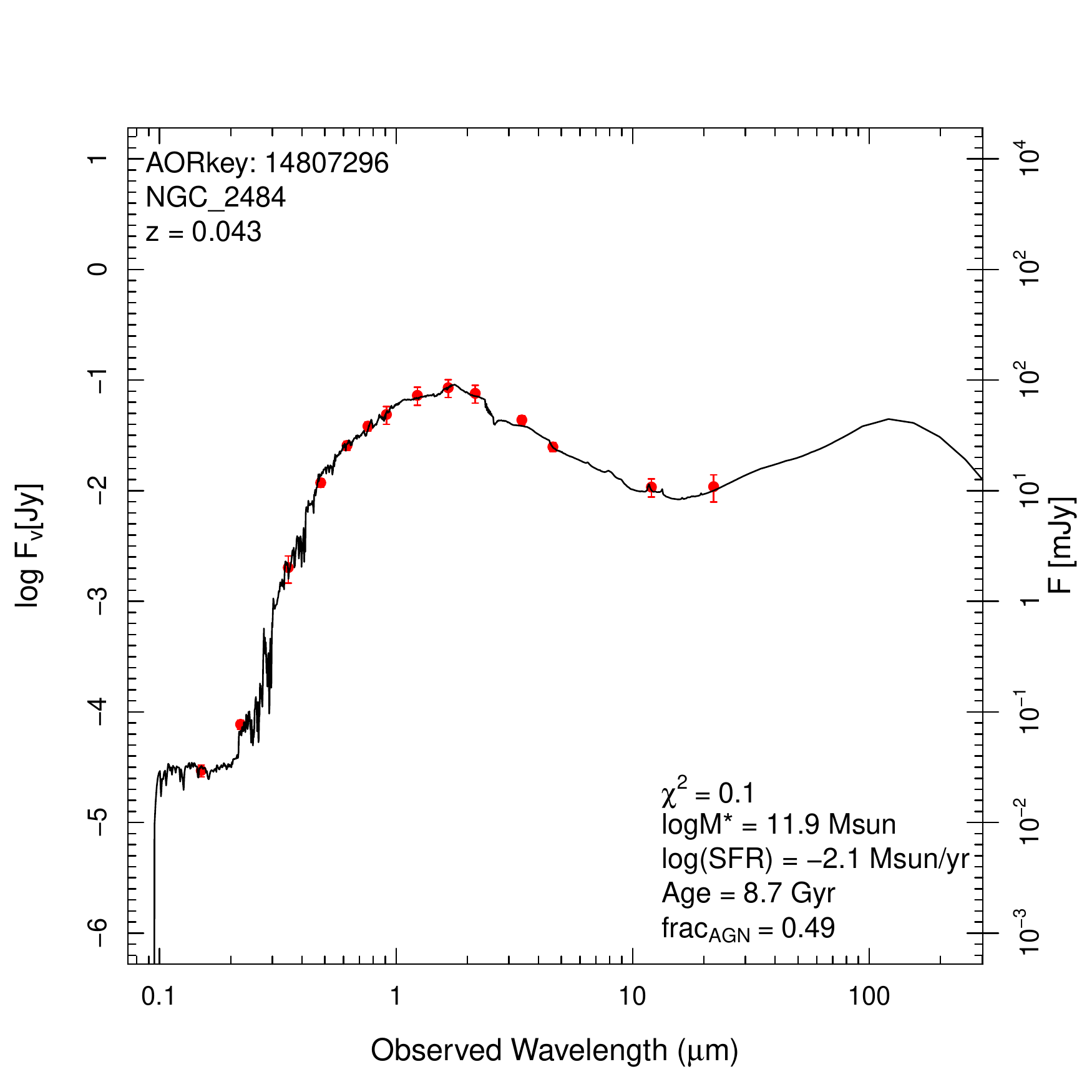}
\includegraphics[height=6cm,width=6cm]{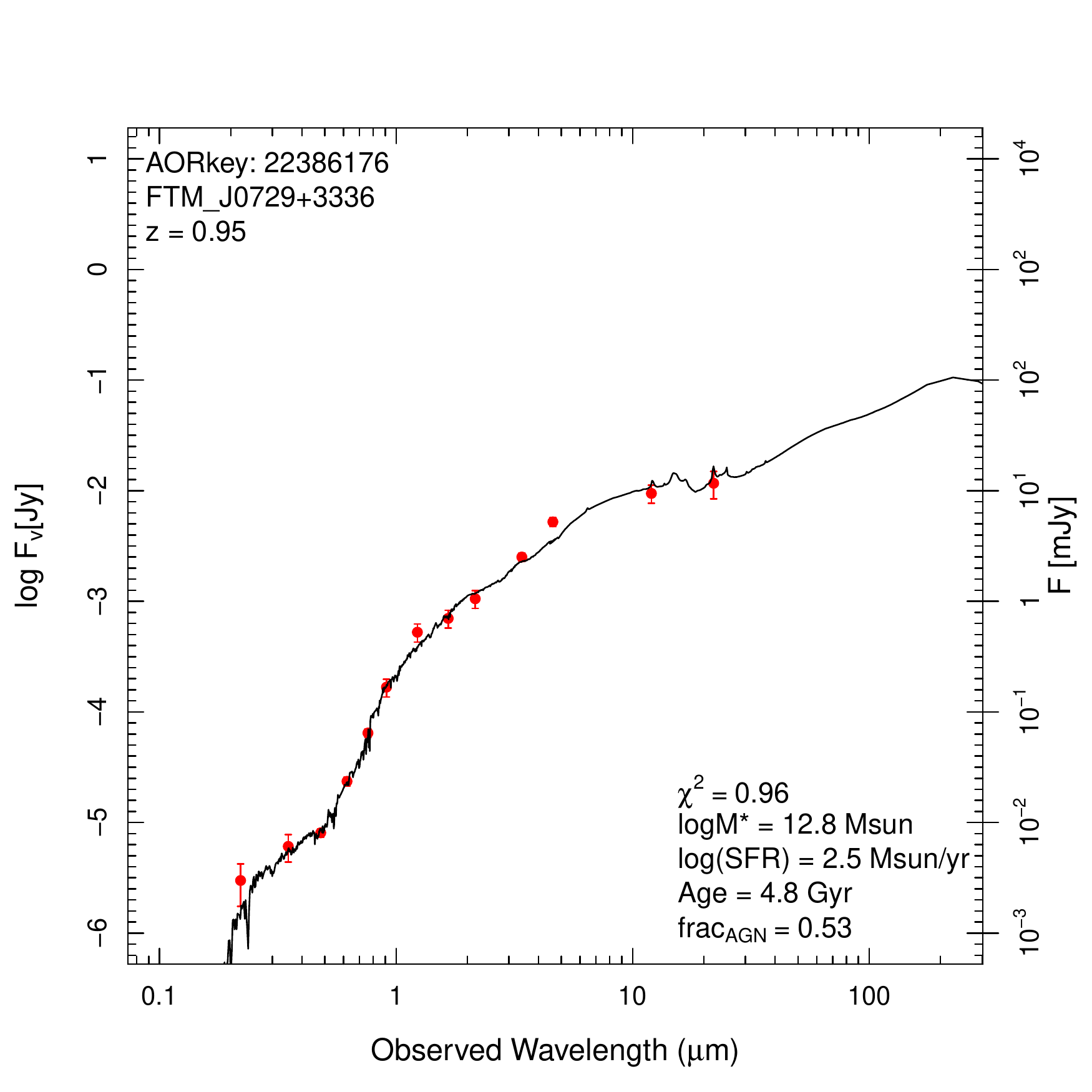}
\includegraphics[height=6cm,width=6cm]{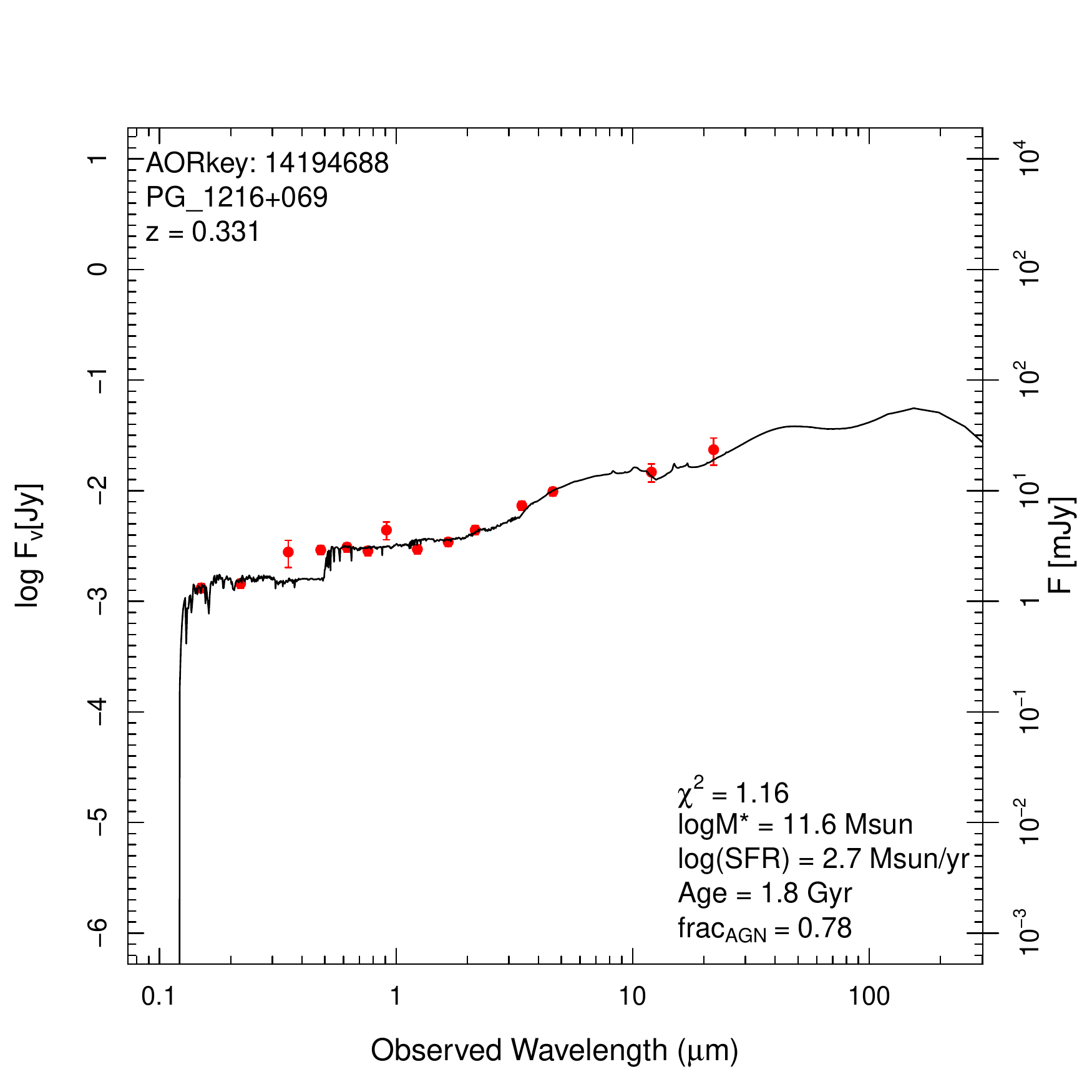}
\includegraphics[height=6cm,width=6cm]{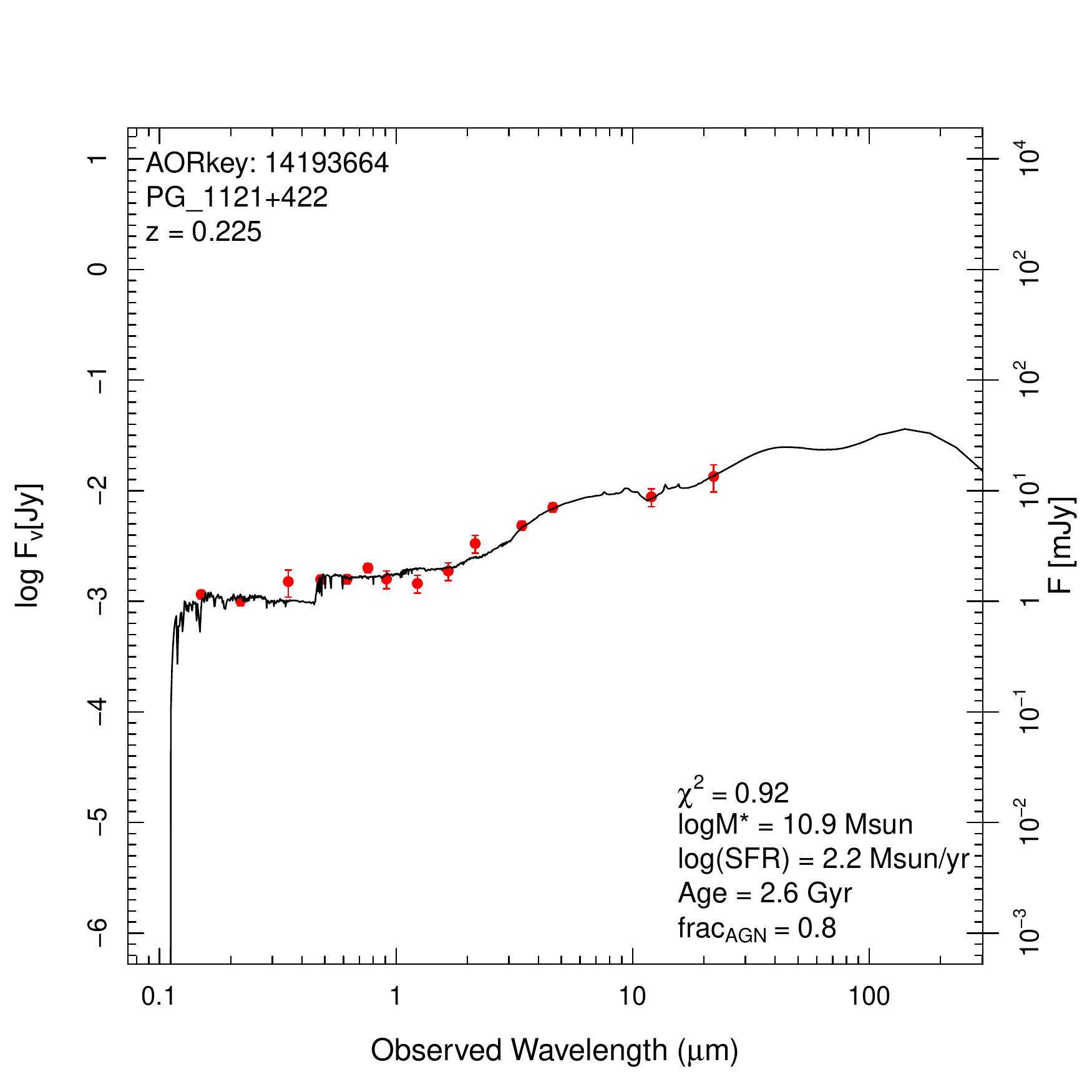}
\caption{Twelve examples of best-fit models chosen to be representative of our sample, arranged with increasing ${\rm frac}_{\rm AGN}$ (top to bottom). 
The observed data are plotted with red points and the best CIGALE model with a solid line.  At the top left of each panel we indicate the Spitzer identification number (AORkey), the NED galaxy name and the redshift. 
The CIGALE output physical parameters together with the minimised chi square of the fit are presented at the bottom right.
}
\label{fig:exmpl1}
\end{figure*}

\section{Measuring the global properties}
\label{sec:results}

\subsection{SED modelling}

We use the Code Investigating GALaxy Emission (CIGALE) software to fit the photometric SED (\citealt[Burgarella et al. in prep, Boquien et al. in prep]{tex:CC15}). 
CIGALE can model the galaxy SEDs from the UV to the sub-mm by assuming a stellar population library and SFH provided by the user. 
The SED is built by taking into account the energy balance, i.e., the energy absorbed by dust in UV-optical and reemitted in the IR, while a Bayesian-like analysis is used to derive the galaxy properties. 
In addition to the dust emission in IR the code can model the AGN emission and estimate its contribution to the overall IR emission.

At first CIGALE computes the unattenuated stellar emission from the stellar population models and SFH. Then the attenuation is computed assuming an attenuation law and the amount or energy attenuated is re-emitted in IR. 
But only then, at the end the AGN emission is added.
 In this paper we use a delayed star-formation history defined as:
\begin{equation}
SFR(t) \propto t\,e^{-t/\tau_{\rm main}}.
\end{equation}
\noindent where $t$ is the time and $\tau_{\rm main}$ the e-folding time of the stellar population.
With this definition, a small value of $\tau_{\rm main}$ such as 1\,Gyr will correspond to a typical early-type galaxy whereas a higher value of 10\,Gyr will provide a more constant SFR with time, typically observed in late-type galaxies.
It has been shown in \citet{tex:CC15} that, in addition to consistently better estimating stellar masses and SFRs, compared to other SFHs (e.g. exponential forms), the delayed SFH provides also a better estimate of the age of the galaxy.
However, we add the possibility for a recent partial quenching or burst of star-formation of specific duration to account for a recent variation in star forming activity of the galaxies.
 Based on the \citet{tex:CB16} founding we used two values, i.e. 0.1 and 0.4 Gyr, for these episodes.
This is handled through the $r_{\rm SFR}$ parameter which is defined as the ratio between the instantaneous SFR and the SFR just before the quenching or starburst.
A value lower than 1 will imply a recent reduction of star forming activity whereas a value higher than 1 will model a starburst.
More details on this parameter can be found in \citet{tex:CB16} and Ciesla et al. (in prep.).
The age corresponds to the age of the oldest stars in the galaxy.
The stellar populations models of \citet{tex:BC03} are then convolved with the delayed SFH to model the non-attenuated stellar emission.

The energy absorbed by the dust, assuming a \citet{tex:CA00} law is re-emitted in the IR using the \citet{tex:DH14b} dust emission templates.
Finally, the code adds the emission from an AGN to the stellar and dust SED.
The AGN templates are from the \citet{tex:FF06} library and build a two component AGN SED.
The first component is an isotropic emission of the point-like central source.
This emission is a composition of power laws with variable indices in the wavelength range of 0.001-20 $\mu$m. 
The second component is radiation from dust with a toroidal geometry close to the central engine.
Part of the direct emission of the AGN is either absorbed by the toroidal obscurer and re-emitted at longer wavelength (1-1000 $\mu$m) or scattered by the same medium.

For each individual galaxy in our sample the modelled flux densities are computed by convolving the modelled SEDs into the set of filters. 
These modelled fluxes are then compared to the observations taking into account the uncertainties on the observed fluxes.
The probability distribution function of each parameter is calculated and the estimated value of the parameter together with the error, corresponding to the mean and standard deviation of this distribution, are derived. 
The CIGALE input parameters, values, and ranges used to create the modelled SEDs are listed in Table \ref{fitparam}.
The initial values for the parameters $\beta,\gamma$ and $\theta$ are kept fixed.
The initial values have been selected based on extensive tests of previous studies, see \citet{tex:CC15} and references therein.
CICALE has been tested for various type of galaxies: local galaxies (\citealt{tex:BG11}), high redshift (z>1) galaxies (\citealt{tex:BH14}) and simulated galaxies that host an AGN component (\citealt{tex:CC15}).
For more details on its current capabilities see \citet{tex:CC15} while for an older version of the model see \citet{tex:NB09}.

\begin{figure}
\centering
\includegraphics[height=12cm,width=9cm]{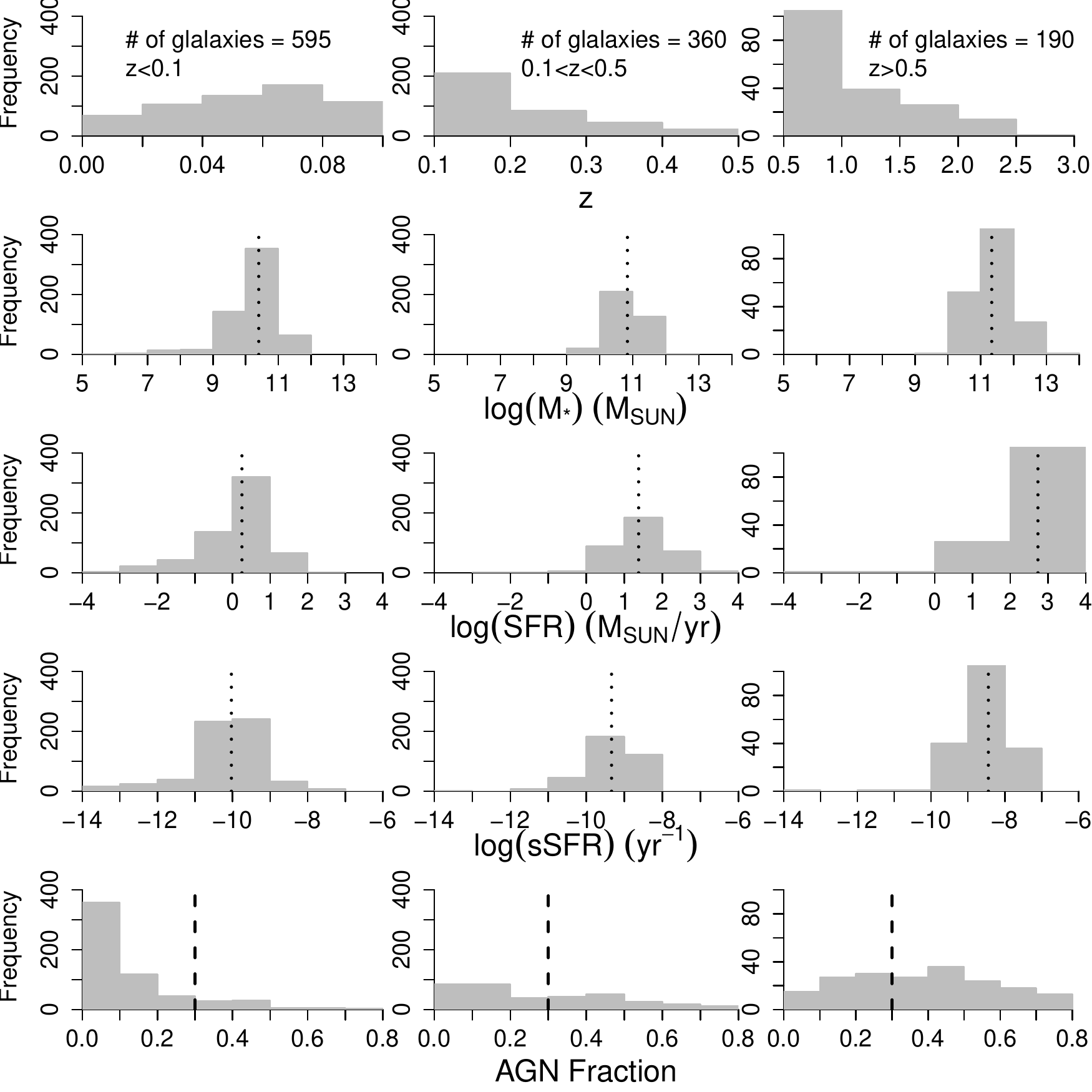}
\caption{Histograms showing the distribution of the global properties of the low redshift (z $<0.1$) galaxies (left panels), the middle redshift ($0.1 \leq \rm z <  0.5$) galaxies (middle panels) and  the high redshift galaxies (z$ \geq 0.5$; right panels). 
From top to bottom: redshift, stellar mass, SFR, sSFR and ${\rm frac}_{\rm AGN}$. 
The dotted vertical lines in the second, third and fourth panel rows indicate the median value of each distribution while the dashed vertical line in the bottom row of panels shows the limit of ${\rm frac}_{\rm AGN}$=0.3.
Below which CIGALE results are not secure enough for obtaining any information about the existence of an AGN.
}
\label{fig:properties}
\end{figure}

\begin{figure}
\centering
\includegraphics[width=0.5\textwidth]{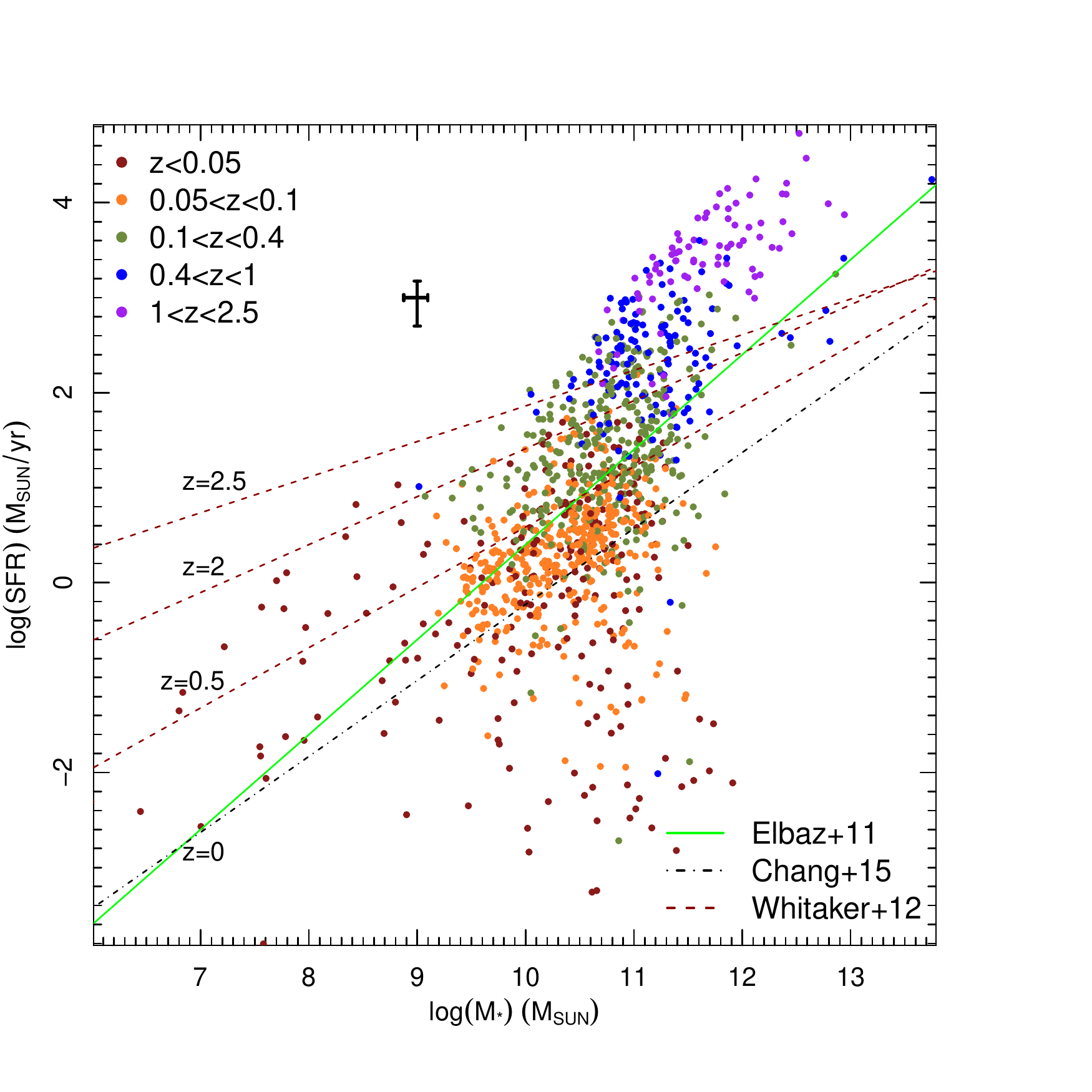}
\caption{The full sample of 1,146 galaxies. 
The colour denotes the redshift bin of each galaxy.
The lines indicate the main sequence (MS) of star-formation as measured by three independent studies at four different redshifts. 
Note that both the black and green lines are at redshift zero.
The error bars show the median uncertainty for the stellar mass and SFR measurements, 25\% and 50\% respectively.
}
\label{fig:mainseq-z}
\end{figure}

\subsection{Basic properties of the sample}

CIGALE estimates various physical parameters for each galaxy, such as the instantaneous SFR (in solar masses per year), 
the stellar mass ($M_\star$, in solar masses), 
the stellar age (in Gyr), 
the AGN luminosity fraction (${\rm frac}_{\rm AGN}$), 
the AGN luminosity (L$_{\rm AGN}$ in W), 
the dust attenuation in the $FUV$ ($A_{\rm FUV}$) 
and the dust luminosity (L$_{\rm dust}$  in W). 
We provide all these physical measurements together with NED cross-matched redshift and galaxy name in Table \ref{table:finalpar}.
The ${\rm frac}_{\rm AGN}$ is the fraction of the total IR luminosity (1-1000$\mu$m) due to the AGN component.
As shown by \citet{tex:CC15}, it is difficult to conclude on the presence of an AGN for galaxies with ${\rm frac}_{\rm AGN}<0.2$ due to high uncertainties, thus in this work we choose to be conservative and only consider as AGNs galaxies with ${\rm frac}_{\rm AGN}>0.3$.

To assess the quality of the results provided by the code, 
we follow the procedure first outlined by \citet{tex:GB11}.
Initially, we run CIGALE on the sample to obtain the best fit SED.
Then, CIGALE convolves these best fit SEDs into the set of filters and add some noise in order to create a mock catalogue of fluxes for each galaxy associated to a known set of output parameters. 
The noise is randomly distributed around the best model flux and within the errors provided by the input photometry.
Finally, we run CIGALE a second time on this mock catalogue and compare the new estimates to the true ones used as input.

In Fig.~\ref{fig:mock} we show the results of this test. 
The stellar mass, SFR, dust luminosity, and AGN luminosity are well constrained by the SED fitting as we see that the estimated values are in a very good agreement with the true ones. 
The estimate of the E(B-V) attenuation is also constrained with our photometric coverage. 
For the ${\rm frac}_{\rm AGN}$ parameter, the results are more dispersed but are consistent with what obtained by \citet{tex:CC15} with high fractions better recovered than smaller ones. 
We note a known effect of the PDF analysis with values close to the maximum (minimum) that tend to be slightly underestimated (overestimated) due to the truncation of the PDF for edge values, for further discussion see \citet{tex:CC15}. 
The age of the oldest stars is not well constrained as expected, showing a flat relation between the estimated values and the true ones.
An alternative parameter that can be used as an age indicator is the sSFR. 
The sSFR can be used to estimate the doubling time of a galaxy, under the assumption that the current SFR has remained constant throughout the lifetime of the galaxy and will do so in the future. 
However, this is a rather strong assumption as it depends strongly not only on the SFH of the galaxy but also on the details of the method to estimate the SFR (\citealt[and references therein]{tex:CW15, tex:BK16, tex:WN16}).

\begin{figure*}
\centering
\includegraphics[height=6cm,width=6.05cm]{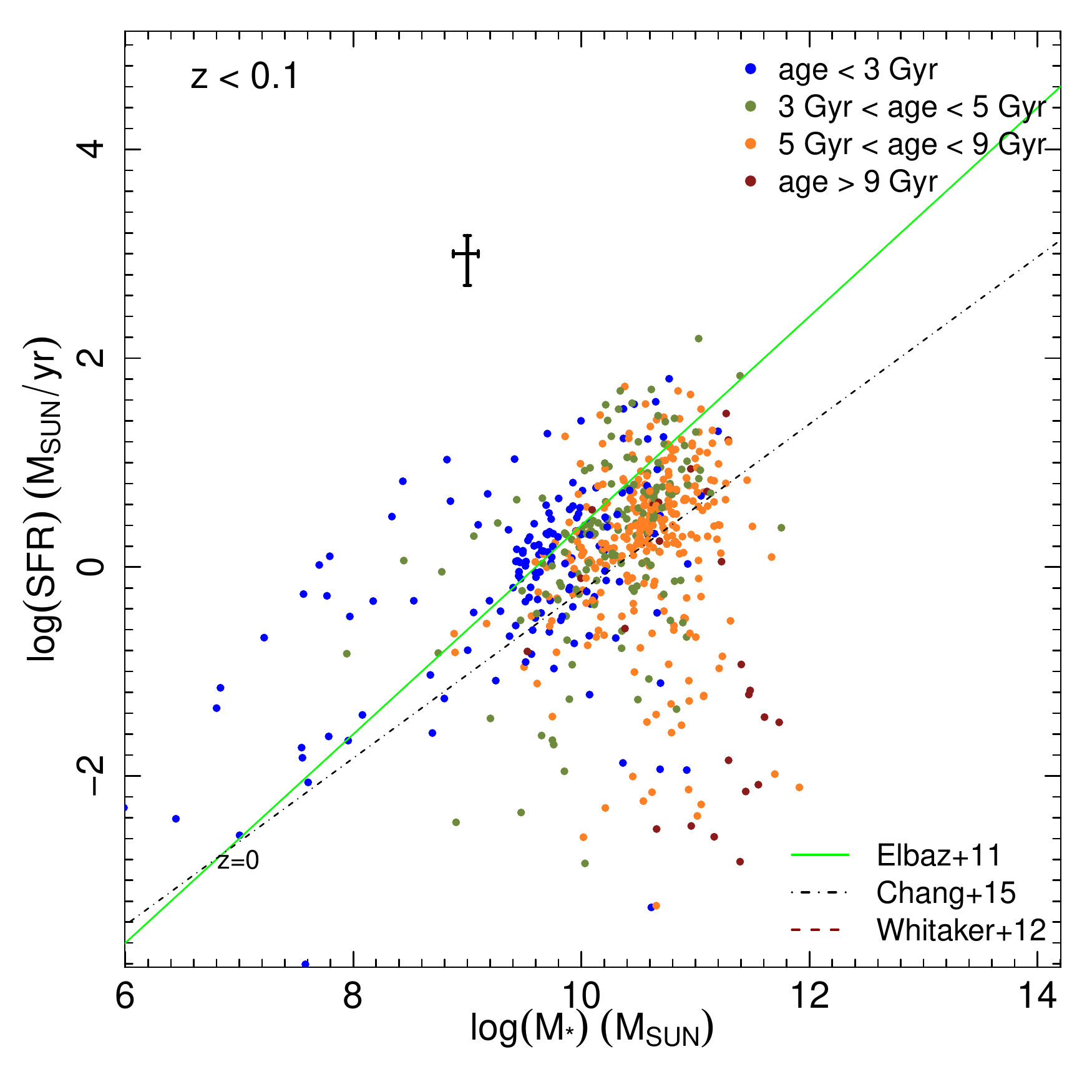}
\includegraphics[height=6cm,width=6.05cm]{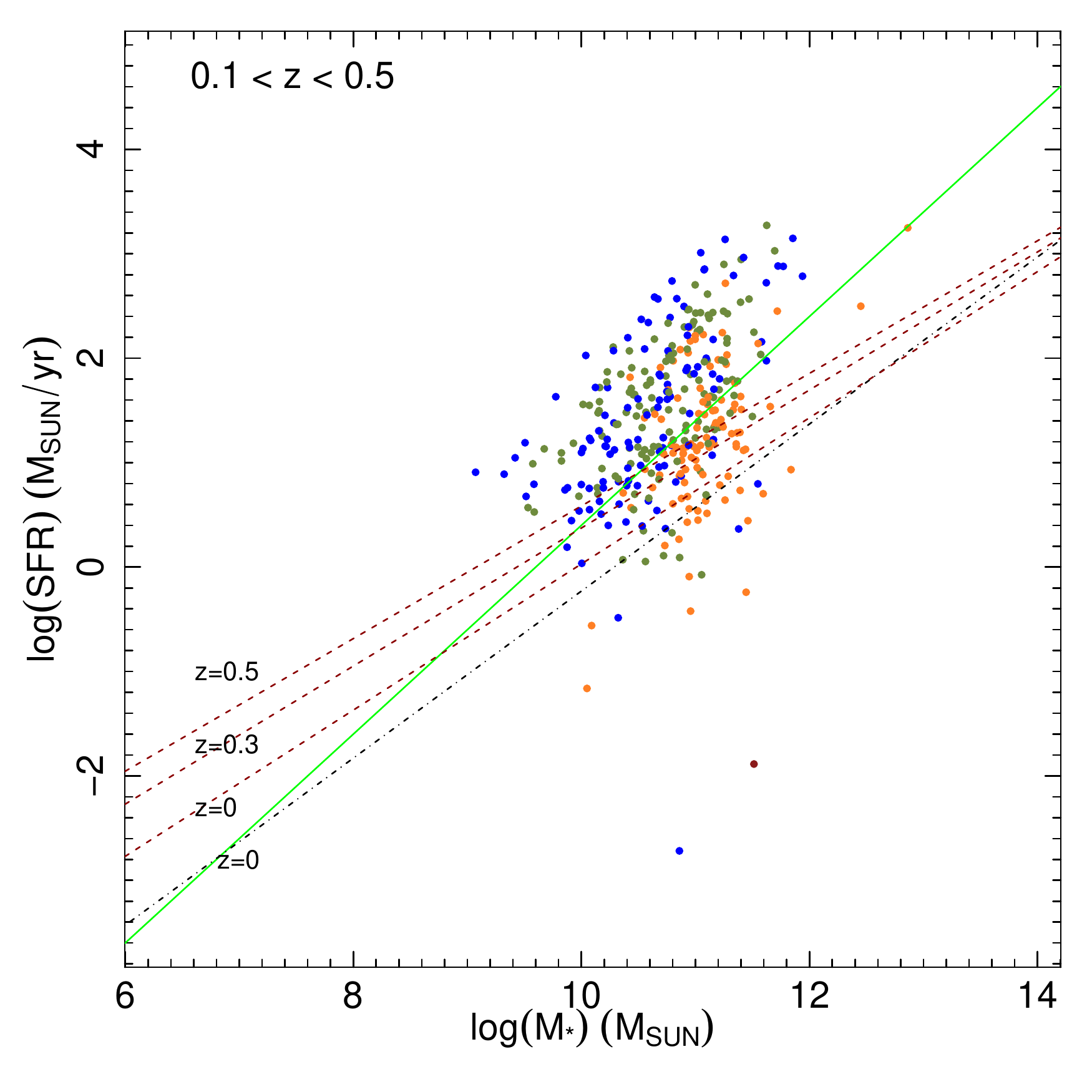}
\includegraphics[height=6cm,width=6.05cm]{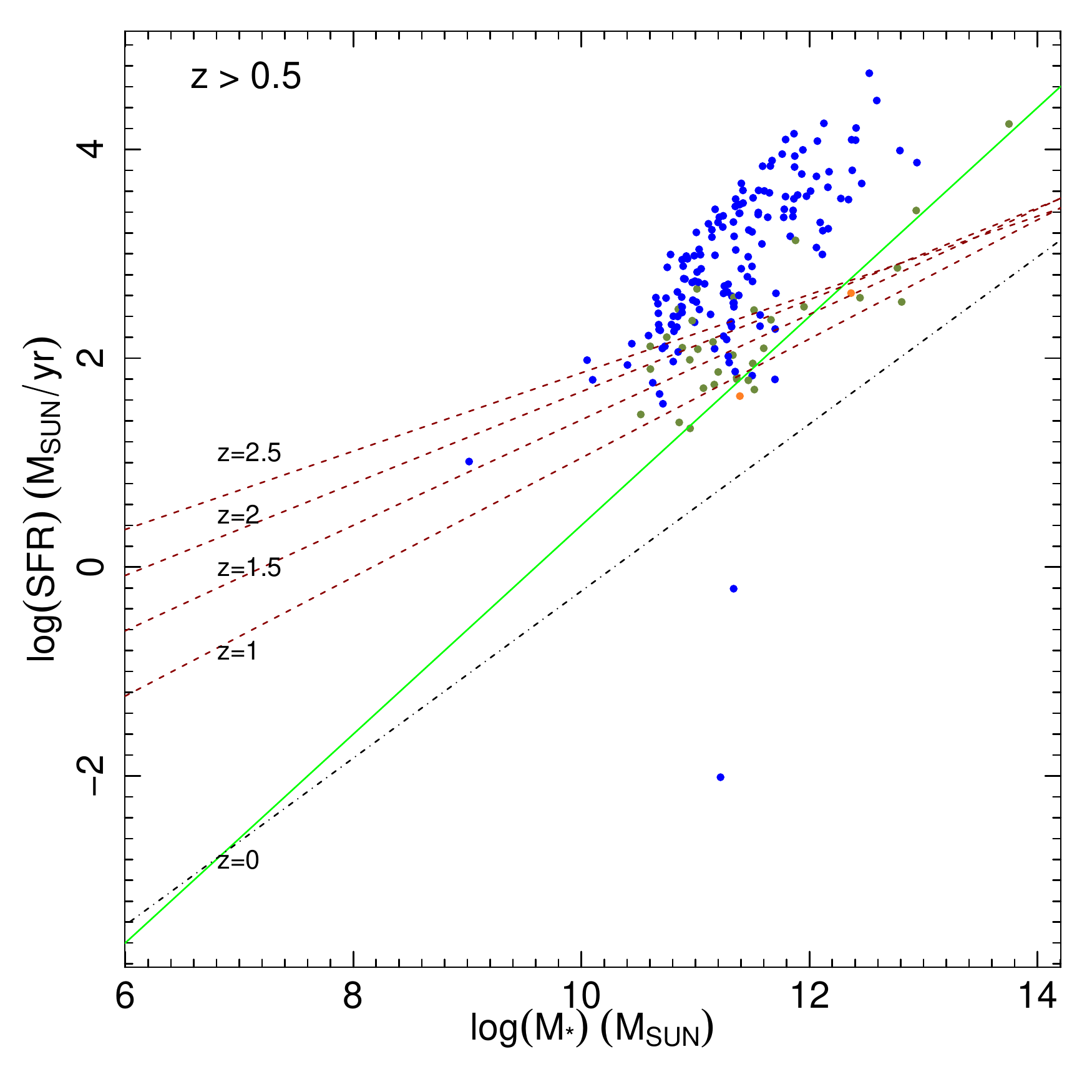}
\caption{The distribution of the 1,146 galaxies in the $\log(\rm SFR)$ - $\log(M_{\star})$ plane. 
The colour coding shows the age of the population where blue colour indicates galaxies with young stellar populations ($<$ 3  Gyr), green colour indicates galaxies that have stellar populations with age between 3-5 Gyr, 
orange colour shows galaxies with stellar population ages between 5 - 9 Gyr and red colour marks galaxies with stellar populations older than 9 Gyr.
Note that the stellar age indicates the oldest stars in a galaxy and does not exclude the existence of younger stellar populations.
Additionally, the stellar age is characterised by high uncertainties and should be treated with caution, see text for more details.
The error bars show the median uncertainty for the stellar mass and SFR measurements, 25\% and 50\% respectively.
The green and black lines, plotted in all the panels, are from \citet{tex:ED11} and \citet{tex:CW15} and represent the SFR$-M_{*}$ relation for the local Universe. 
In the middle panel the red dashed lines represent the MS at redshifts 0, 0.3 and 0.5, while in the right panel at redshifts 1, 1.5, 2, 2.5. 
The red lines are from \citet{tex:WV12}.}
\label{fig:mainseq}
\end{figure*}

\begin{figure*}
\centering
\includegraphics[height=6cm,width=6cm]{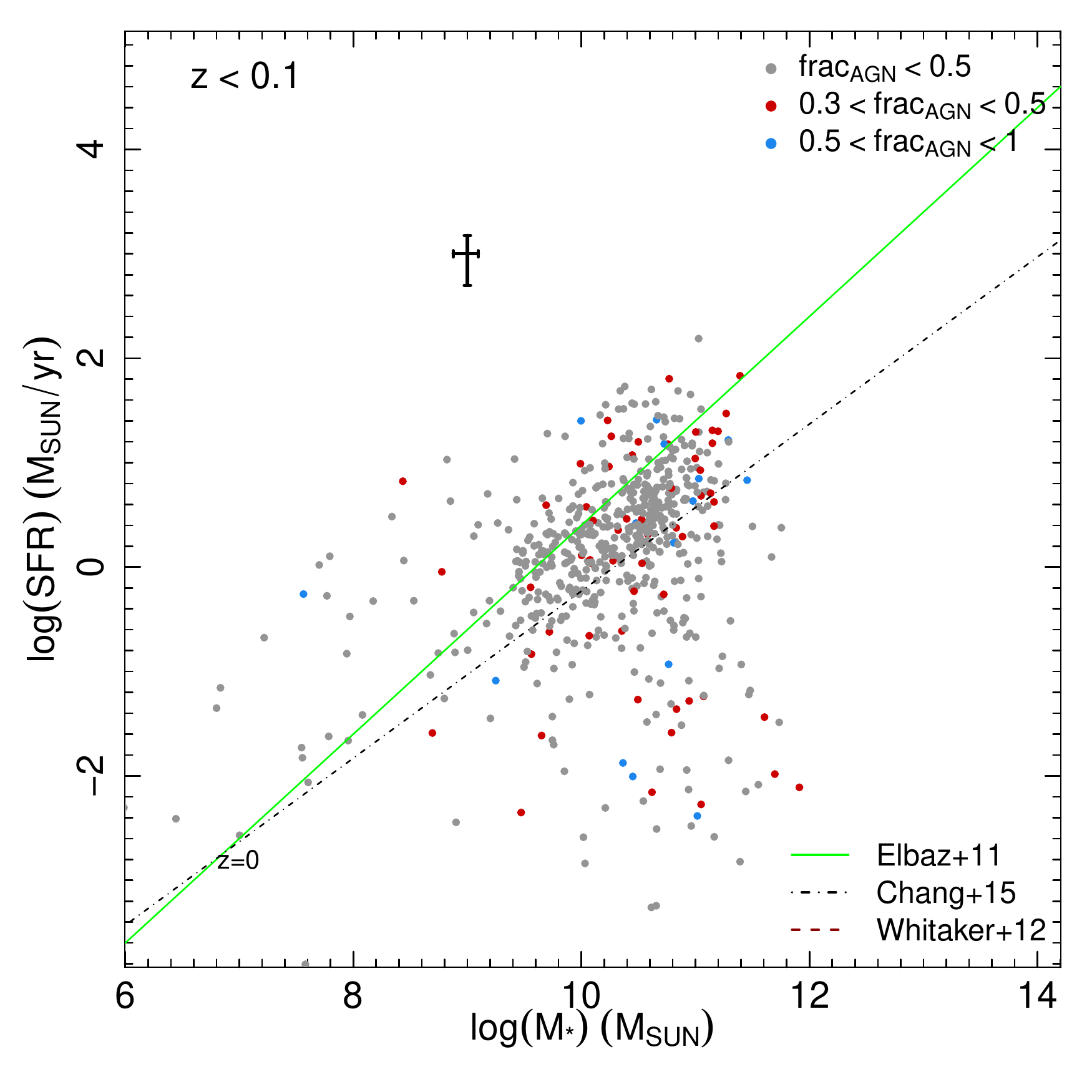}
\includegraphics[height=6cm,width=6cm]{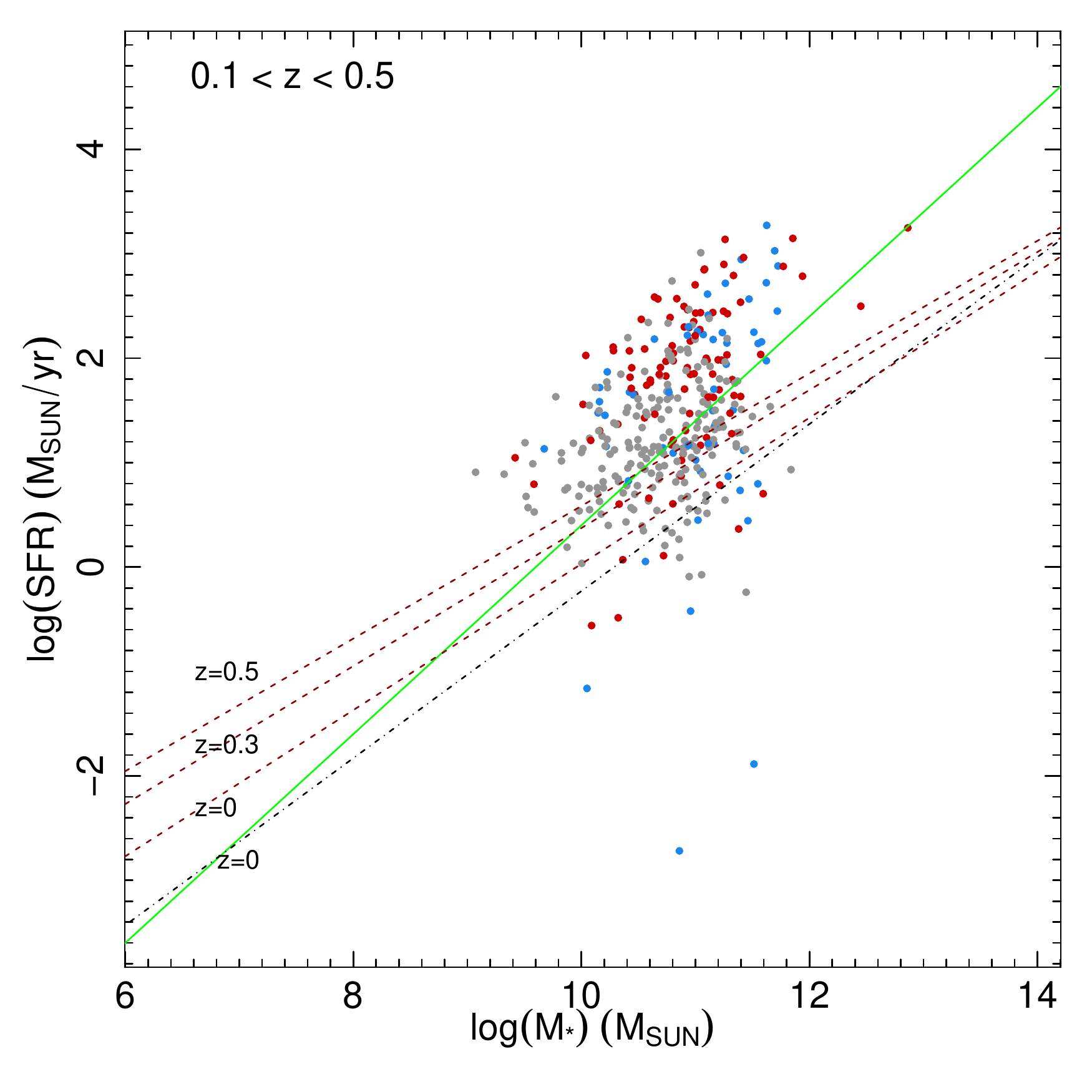}
\includegraphics[height=6cm,width=6cm]{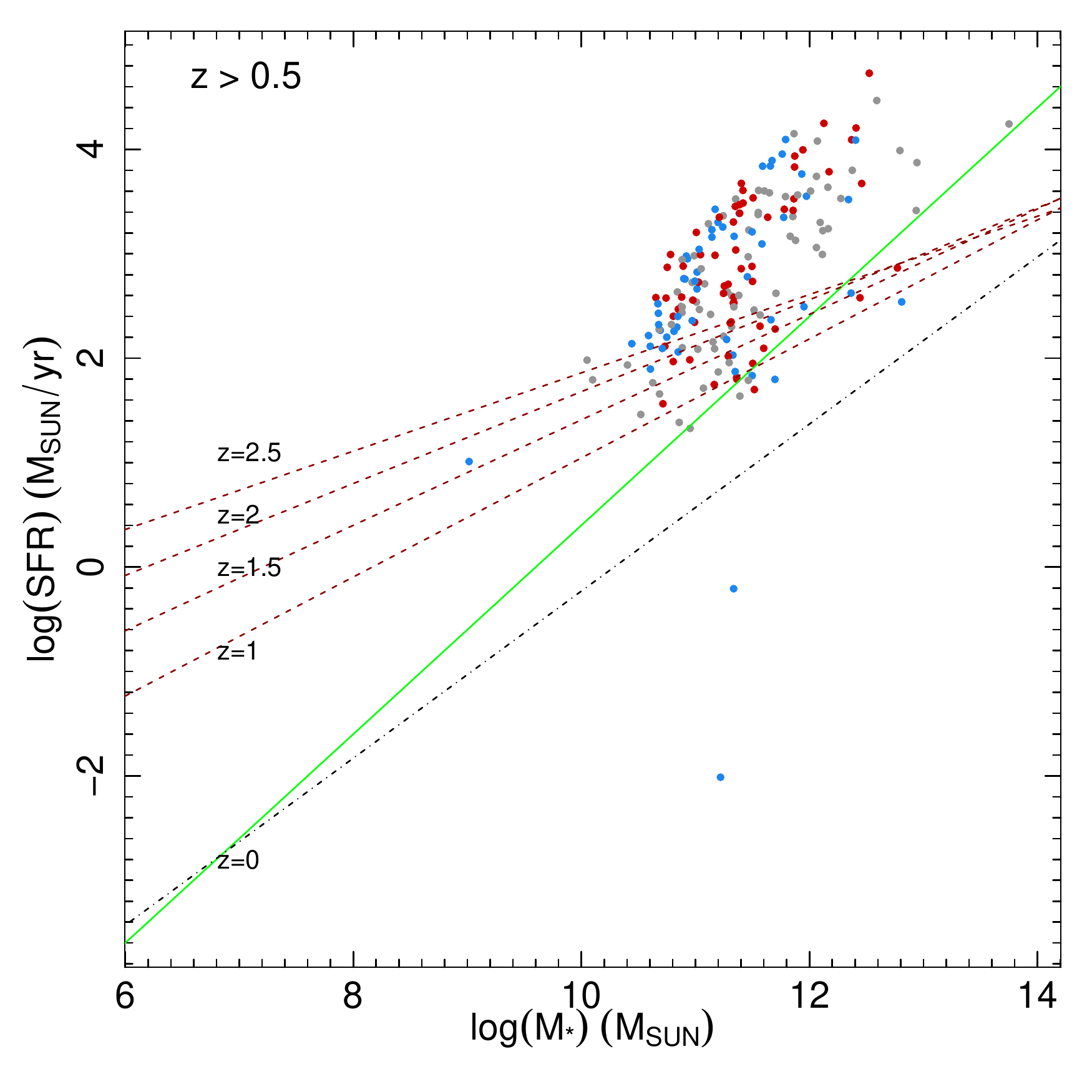}
\caption{The general layout is the same as in Fig.~\ref{fig:mainseq} but here the colour code is based on ${\rm frac}_{\rm AGN}$.
Grey points indicate no AGN contribution, red moderate and blue high AGN contribution to the global IR emission of the galaxy.
}
\label{fig:mainseq-agnfr}
\end{figure*}

As \citet{tex:CB14} and \citet{tex:BH14} showed, a poor infrared coverage does not significantly influence our ability to accurately measure the dust luminosity and SFR with CIGALE as long as one observation in the mid-IR is available.
However, all parameters directly related to SFH (such as stellar age), are challenging to constrain as we can see in Fig.~\ref{fig:mock} and their reported values should be treated with caution.

We wish to inspect if the produced SED models are in agreement with the observed fluxes, since this first verification will support that CIGALE provides realistic physical parameters. 
In additional to the reported minimised chi square ($\chi^2$), we visually examine all the SED models produced for the 1,146 galaxies to ensure that there is a good overall agreement between the CIGALE model and the observed flux densities of each galaxy. 
In Fig.~\ref{fig:exmpl1} we illustrate twelve such examples at different redshifts.

In Fig.~\ref{fig:properties} we show the distribution of redshift, stellar mass, SFR, sSFR and ${\rm frac}_{\rm AGN}$ for the full sample. 
To facilitate the current study we split the sample in three subsamples based on redshift. 
The first sample contains 584 galaxies with z$<0.1$, the second sample contains 360 galaxies with redshift $0.1\leq \rm z<0.5$ while the third group contains the remaining 190 galaxies with z$ \geq 0.5$. 
Fig.~\ref{fig:properties} shows a gradual increase with redshift of the stellar mass, SFR and sSFR median value. 
This is a result both of a selection bias, i.e. the sample contains more massive and luminous systems at higher redshifts, and a physical result, 
i.e. galaxies of the same stellar mass having a higher SFR at higher redshifts as it has been reported by other studies (e.g. \citealt{tex:DD07}).
 
As we mentioned earlier, our sample is rather diverse and spans a large range of redshifts, $M_{\star}$ and SFRs, mainly due to the different criteria used to target the original galaxies for Spitzer/IRS spectroscopy.
Figure~\ref{fig:mainseq-z} illustrates the connection between the stellar mass, SFR and redshift for the full sample of 1,146 galaxies.
Galaxies with z$<$0.05 (218 galaxies) span from dwarfs (M$_{\star}\sim10^{6}$ M$_{\odot}$) to very massive galaxies (M$_{\star}\sim10^{12}$ M$_{\odot}$). 
Their locus in the SFR$-M_{*}$ plane extends from non star-forming galaxies well below the main sequence (MS) to highly star-forming galaxies with sSFR of $\sim 10^{-8}$ yr$^{-1}$.
Only in Figure~\ref{fig:mainseq-z} we use more redshift bins in order to better highlight how the SFR and stellar mass are changing with redshift.
Galaxies with redshift 0.05$\leq$z$<$0.1 (378 galaxies) are mainly on the corresponding MS of star-formation. 
While the 550 galaxies at 0.1$\leq$z$<$2.5, are above the MS corresponding at each redshift, indicating that the Spitzer/IRS selection was indeed targeting extremely luminous, star forming galaxies.

From Figures~\ref{fig:properties} and \ref{fig:mainseq-z} it is easily seen that the Spitzer/IRS, CASSIS catalogue does not provide a complete sample of galaxies. 
However, this sample has the great advantage to contain all the galaxies that have mid-IR spectrum observations and broadband photometry from UV to 22$\mu$m. 
The analysis presented in this paper will allow future studies to explore connections between the global physical properties presented here and the spectroscopic properties derived from IRS spectrum.

\section{Exploring the physical properties of CASSIS galaxies}  
\label{sec:phypro}

In Figure~\ref{fig:mainseq-z} we showed the distribution of the full sample on the $\log (\rm SFR) - \log (M_{\star})$ plane. 
In Figures~\ref{fig:mainseq} and \ref{fig:mainseq-agnfr} we split the sample in the same three sub-samples namely z$<0.1$, 0.1$\leq $z$<$0.5 and z$\geq $0.5, as in Fig.~\ref{fig:properties}. 
In all three Figures the local MS of star forming galaxies is shown as measured from two independent studies, \citet{tex:ED11,tex:CW15} in green and black respectively.
Additionally to the local MS the z=1, 1.5, 2 and 2.5 lines are also plotted \citet{tex:WV12}.

In Fig.~\ref{fig:mainseq} we colour code our galaxies based on their stellar age as measured by CIGALE.
In the left panel (z$<0.1$) we see a gradual increase of the stellar age as we move to more massive galaxies with lower star-formation. 
At the top left the dwarf galaxies with very young stellar populations are reside well above the MS, while as we move to the bottom right we find mainly galaxies with age above 9 Gyr.
Galaxies below the MS are believed to be red galaxies with quenched star-formation (\citealt{tex:WV12}).
The dispersion of old age galaxies (red and orange points) is not unexpected as the age parameter indicates the oldest star in a galaxy. 
Thus, it is possible that a galaxy contains a very old stellar population and at the same time has high star-formation rate, since its age is not related to the current SFR.
It is important to remind the reader that CIGALE stellar age is not well constrained as we showed in Section 3.2 and should be treated with caution.

It has been found that there is an evolution on the MS relation, with galaxies at higher redshifts forming stars at higher rates compared to galaxies of the same stellar masses at lower redshifts (e.g. \citealt{tex:DD07}).
Our galaxies in the middle ($0.1 \leq \rm z<0.5$) and right (z$ \geq 0.5$) panels of Fig.~\ref{fig:mainseq} are always populating the area above the MS corresponding to the relevant redshift bin.
For a more detailed comparison between each galaxy redshift and the placement on the $\log (\rm SFR) - \log (M_{\star})$ plane see Fig.~\ref{fig:mainseq-z}.
It is unclear what is the nature of these galaxies that are found above the MS which are the vast majority of mid and high-redshift galaxies in our sample. 
Some studies characterise them as dusty, with blue colour and AGN activity (\citealt{tex:WV12,tex:SM15b}).
However, other studies have shown that AGN can populate both the area on  (\citealt{tex:MP12}) and below (\citealt{tex:MA15c}) the MS.
In the next paragraph and section we try to find out what type of galaxies are contained in the CASSIS sample and where they reside on the MS using both the CIGALE outputs together with alternative measurements found in literature.

\begin{figure}
\centering
\includegraphics[width=0.5\textwidth]{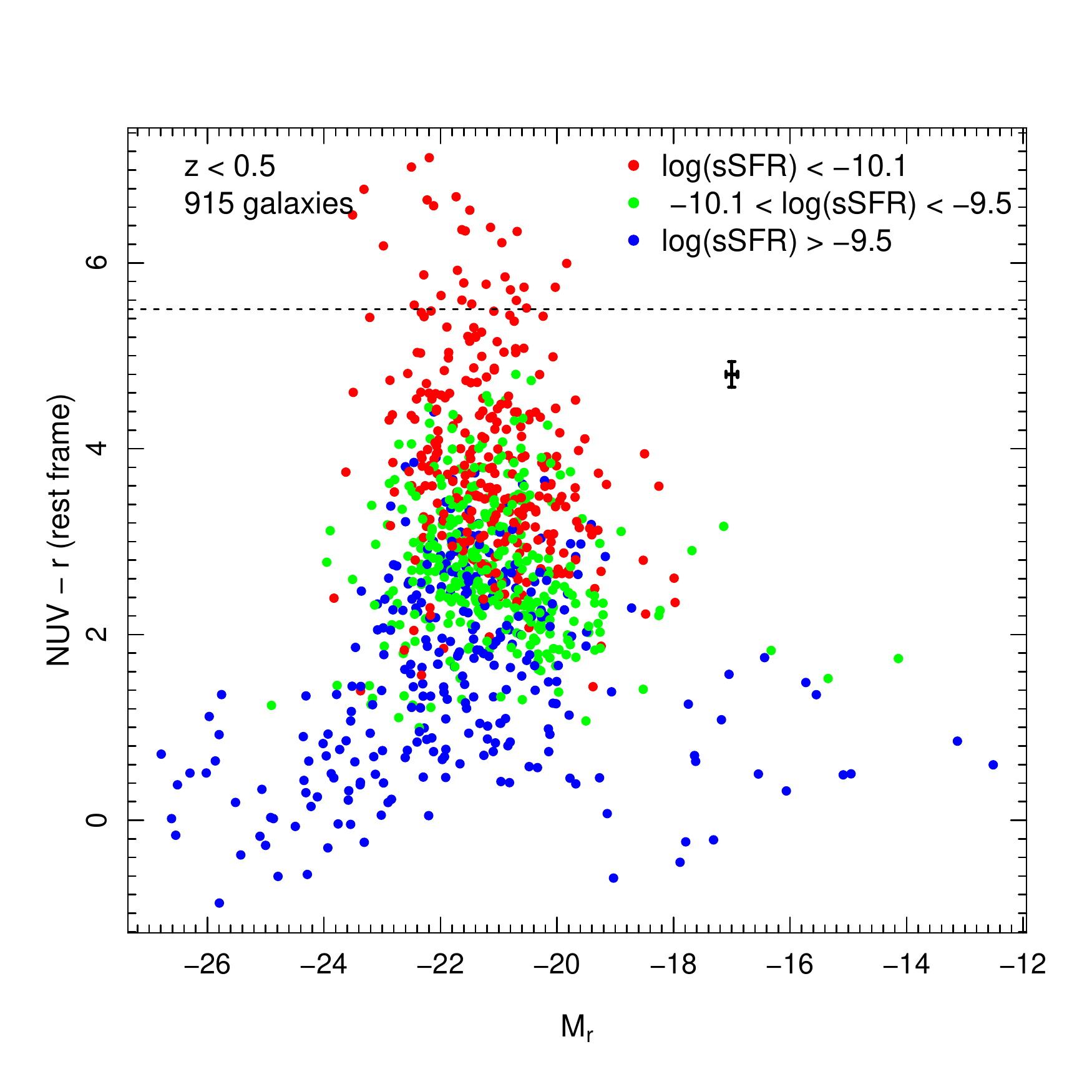}
\caption{The colour magnitude relation for a subsample of 915 galaxies. 
The rest frame magnitudes were calculated using K-corrections calculator (\citealt{tex:CZ12}). 
The distances used for calculating the $r-$band absolute magnitude (M$_{\rm r}$) are based on distances from the literature.
The colour coding is based on the sSFR of each galaxy.
The dashed line indicates the empirical separation between galaxies with recent episode of star-formation (below the line) and quenched galaxies (above the line). See text for more details.}
\label{fig:colmag_ssfr}
\end{figure}

Thus, based on the ${\rm frac}_{\rm AGN}$ CIGALE parameter we examine how star-forming and AGN galaxies are distributed in the $\log (\rm SFR) - \log (M_{\star})$ plane.
Figure~\ref{fig:mainseq-agnfr} is similar to Fig.~\ref{fig:mainseq} but this time the colour of each galaxy corresponds to the ${\rm frac}_{\rm AGN}$. 
In the left panel (z$<0.1$) we find that the majority of the galaxies have ${\rm frac}_{\rm AGN}$ smaller than 0.3 while 10\% and 2\% have moderate and high AGN contribution.
In the middle panel ($0.1 \leq \rm z<0.5$) we notice an increase of galaxies with moderate and high AGN contribution on the IR emission to 18\% and 13\% respectively. 
While, in the right panel (z$ \geq 0.5$) the percentage of AGN galaxies is even further increased to more than half of the sample, 29\% and 24\% respectively.
Finally, we find that CASSIS AGN galaxies are not found on a specific area on the SFR$-M_{\star}$ plane but instead are distributed across all the stellar masses and SFRs in all three panels and can be found above, on and below the MS.

\subsection{Colour and structural properties of the low- and mid- redshift sub-samples}  

In order to expand the investigation of our sample we employ alternative measurements from other studies.
In this section we will only study galaxies with available $NUV-r$ rest frame colour and structural measurements.
In the first case the sample is limited to 915 galaxies with redshift z $<0.5$ and in the second case to 256 galaxies with z $<0.3$.
All galaxies of the second group are included in the first group.

\begin{figure}
\centering
\includegraphics[width=0.5\textwidth]{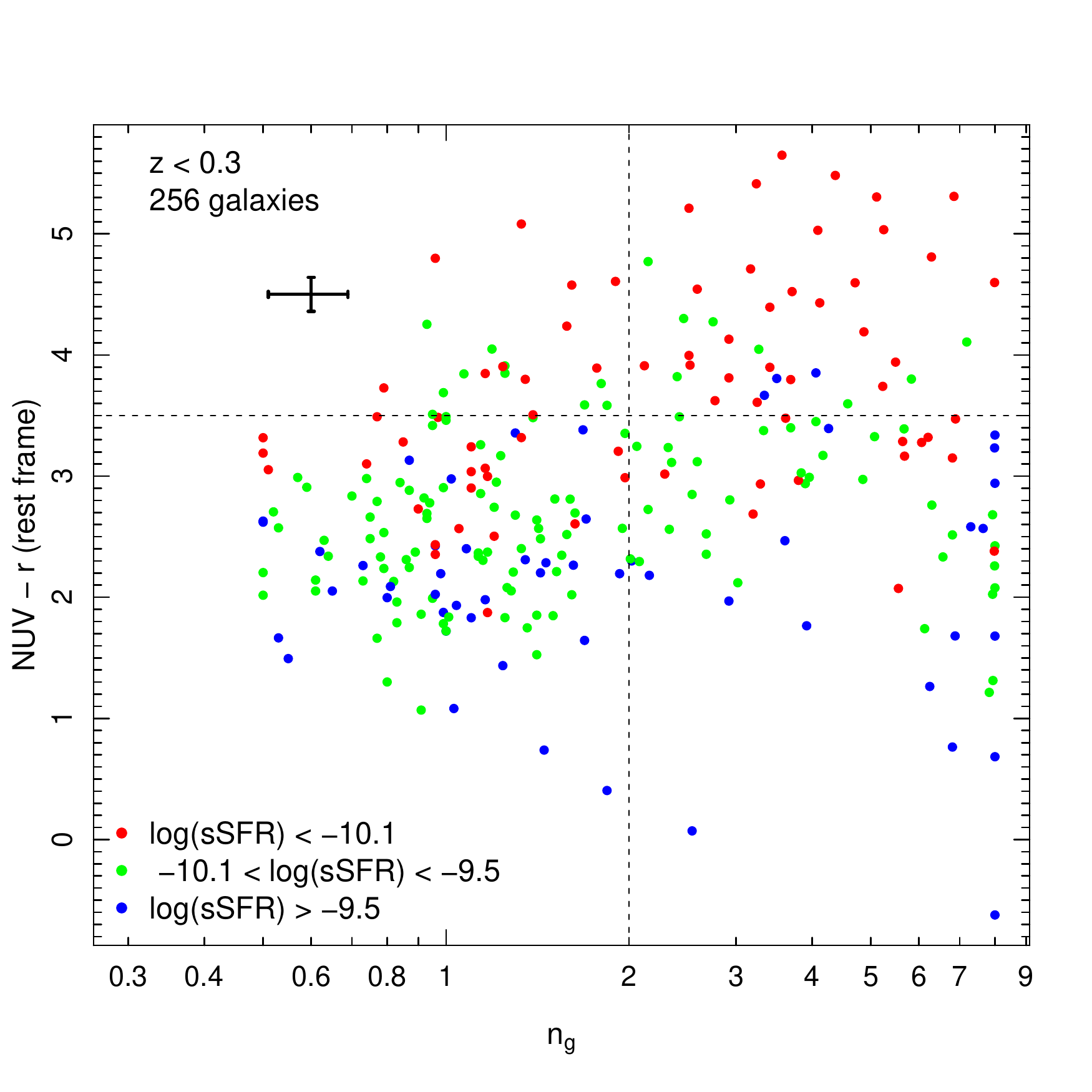}
\caption{Distribution in the ($NUV-r$) - $n_{\rm g}$ plane of 256 galaxies that have structural information.
The rest frame magnitudes were calculated with the use of K-corrections calculator (\citealt{tex:CZ12}). 
The single $g-$band S\'ersic index ($n_{\rm g}$) is from \citet{tex:SM11}.}
\label{fig:colour-n}
\end{figure}

First, we explore the z $<0.5$ subsample with available $NUV-r$ colours by employing the colour magnitude diagram. 
The colour distribution of local galaxies in this diagram is nearly bimodal and relates to galaxy morphology. 
In Fig.~\ref{fig:colmag_ssfr} we plot the $NUV-r$ colour $r-$band absolute magnitude diagram. 
Ultraviolet bands are excellent tracers of the recent star-formation and the $NUV-r$ colour diagram allows to separate the red, non-star forming galaxies from the blue, star-forming galaxies. 
The dashed line is an empirical distinction between galaxies with recent episode of star-formation ($NUV-r<5.4$) and quenched galaxies (\citealt{tex:SI01,tex:SK07}).

We note that our sample lacks quenched galaxies and the vast majority of the galaxies have experienced some recent star-formation. 
This is expected since most galaxies had to be mid-IR bright (late type/LIRGs/QSO) to be targeted by IRS.
We colour-code our galaxies based on the sSFR and we find that there is a clear correlation between CIGALE measured sSFR and the $NUV-r$ colour, where sSFR decreases as the galaxy is becoming redder.
This finding supports the empirical use of this diagram to distinguish between star-forming and non star-forming galaxies.


In order to investigate the structure of the galaxies in our sample we retrieve $g-$band S\'ersic index ($n_{\rm g}$) measurements from the literature. 
The largest catalogue that provides structural measurement is based on local galaxies, z$<0.3$, of the SDSS catalogue (\citealt{tex:SM11}).  
In Fig.~\ref{fig:colour-n} we use the $NUV-r$ colour - S\'ersic index plane to separate between early (E, S0, Sa) and late-type galaxies (Sa-Irr). 
The addition of the S\'ersic index cut makes the separation more effective than using the absolute magnitude or a single colour cut. 
This is primarily because the S\'ersic index helps us to overcome the colour confusion between old stellar populations in spheroidal systems and edge-on dust attenuated spirals (\citealt{tex:DR12}).

In Fig.~\ref{fig:colour-n}, early-type galaxies reside on the top-right ($n_{\rm g}>2$, $NUV-r>3.5$) while late-type galaxies at the bottom-left ($n_{\rm g}<2$, $NUV-r < 3.5$).
Both red \& low-$n$ ($n_{\rm g}<2$, $NUV-r>3.5$) galaxies and blue \& high-$n$ ($n_{\rm g}>2$, $NUV-r<3.5$) are a mixture of various morphological classifications.
We colour code the sample based on the sSFR.
Early-type galaxies are mainly (72\%) low-sSFR galaxies and 20\% are mid-sSFR galaxies. Late-type galaxies are mainly mid-sSFR (58\%) while a 25\% are high-sSFR and a 17\% are of low-sSFR. 
We also find that red \& low-$n$ galaxies  are low-, mid-sSFR galaxies while blue \& high-$n$ galaxies are a mixture of all three sSFR bins.

\begin{figure}
\centering
\includegraphics[width=0.5\textwidth]{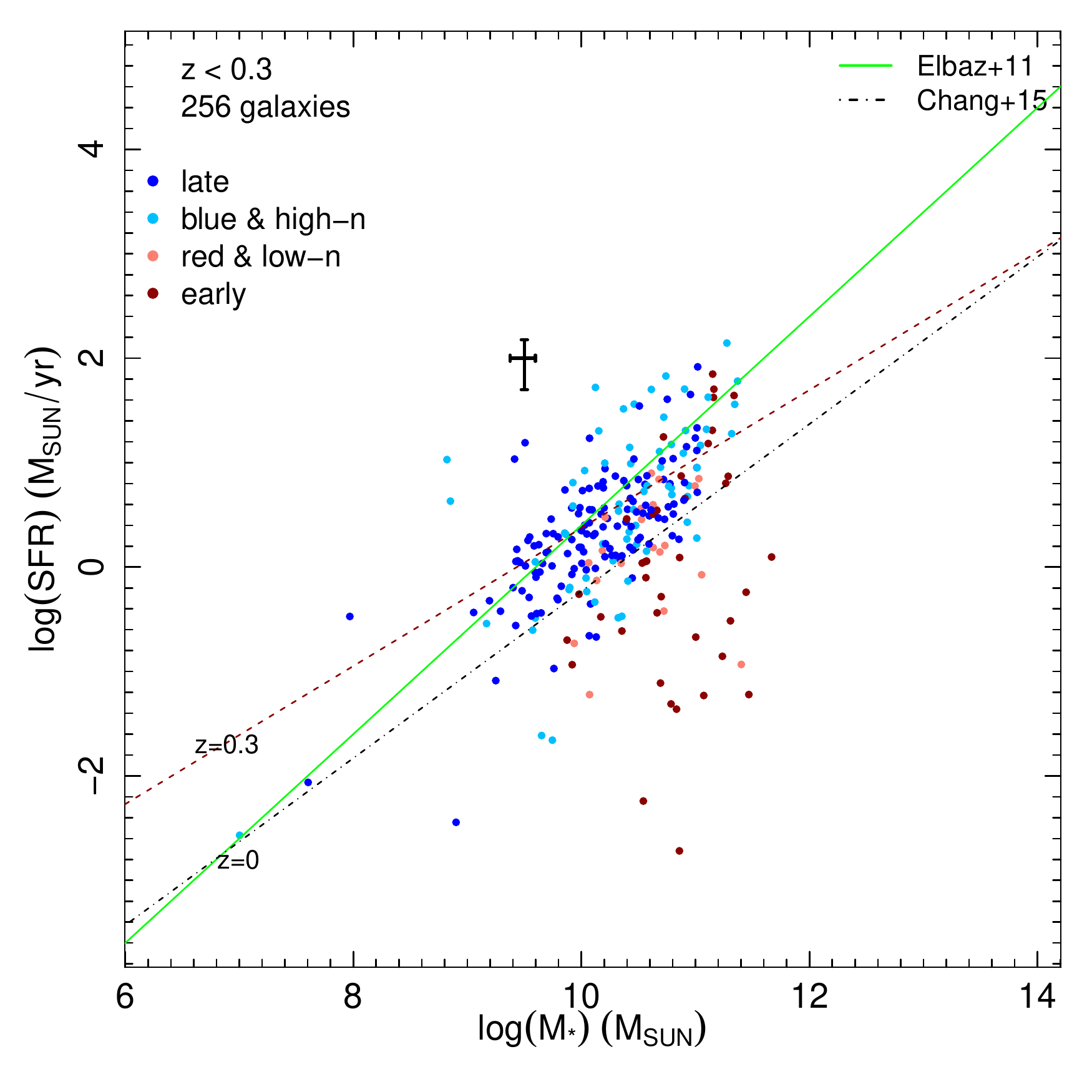}
\caption{ The distribution of the 256 galaxies with structural information in the $\log (\rm SFR) - \log (M_{\star})$ plane. 
The colour coding is based on the galaxy classification from Fig.~\ref{fig:colour-n}. See text for more details.}
\label{fig:mainseq-struc}
\end{figure}

Figure~\ref{fig:mainseq-struc} shows the distribution of the 256 galaxies that are classified based on their position on Fig.~\ref{fig:colour-n}.  
Based on the four quarters defined in Fig.~\ref{fig:colour-n} we have separated the sample in late, blue \& high-n, red \& low-n and early-type galaxies. 
We find a gradual change from blue \& high-n to early-type galaxies  in the $\log (\rm SFR) - \log (M_{\star})$ plane.
Where blue \& high-$n$ galaxies are populating the area above and on the MS showing an extra star-formation activity compared to the late-type population that are found mainly on the MS.
We would like to point out that high-$n$ galaxies are not necessarily bulge dominated galaxies but can host a small bulge that has a pointy central profile that force the S\'ersic index to increase in order to model this extra light in the center.
Considering also that \citet{tex:ED11} found that galaxies placed above the MS are nuclear starburst galaxies we can speculate that blue \& high-$n$ galaxies are late-type starburst galaxies with strong unresolved nuclear emission that leads to high values of $n$.
Galaxies that are classified as red \& low-$n$ are also mainly star-forming galaxies while early late type are either non-star forming galaxies or star forming but slightly below the late-type galaxies.

\begin{figure}
\centering
\includegraphics[width=0.5\textwidth]{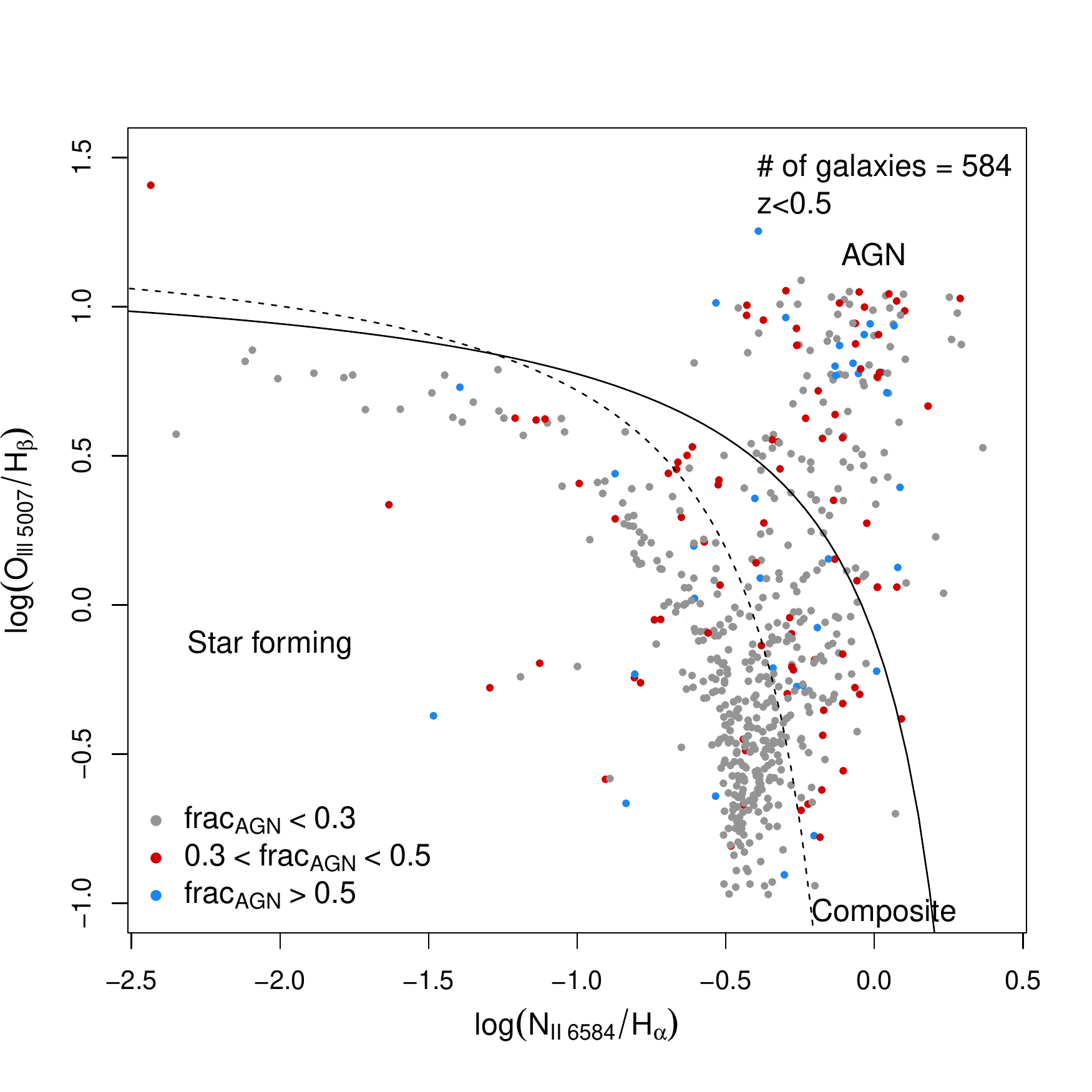}
\includegraphics[width=0.5\textwidth]{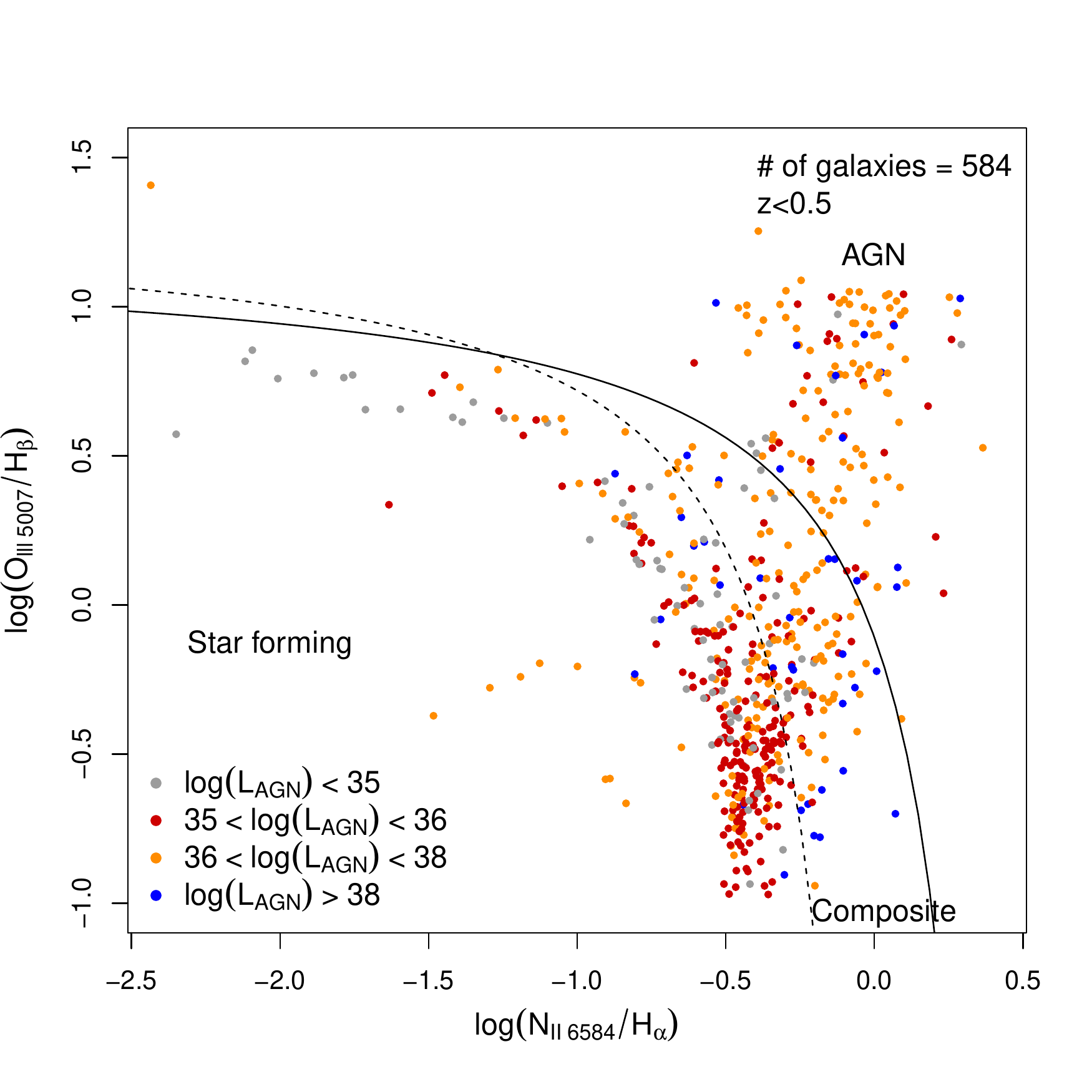}
\caption{The BPT line diagnostic diagrams for all galaxies with z$<0.5$ that spectral lines measurements from SDSS exist. 
The solid line is from  \citet{tex:KG06} while the dashed line from \citet{tex:KH03c}.
The colour coding on the top panel indicates the ${\rm frac}_{\rm AGN}$ and at the bottom panel the AGN luminosity as measured by CIGALE.
The median line flux uncertainty is 3\% for the [O$_{\rm III}$], 2\% for the [N$_{\rm II}$], 3\% for the  [H$_{\beta}$] and 1\% for the [H$_{\alpha}$]. }
\label{fig:bpt-agnfr}
\end{figure}

\begin{figure}
\centering
\includegraphics[width=0.5\textwidth]{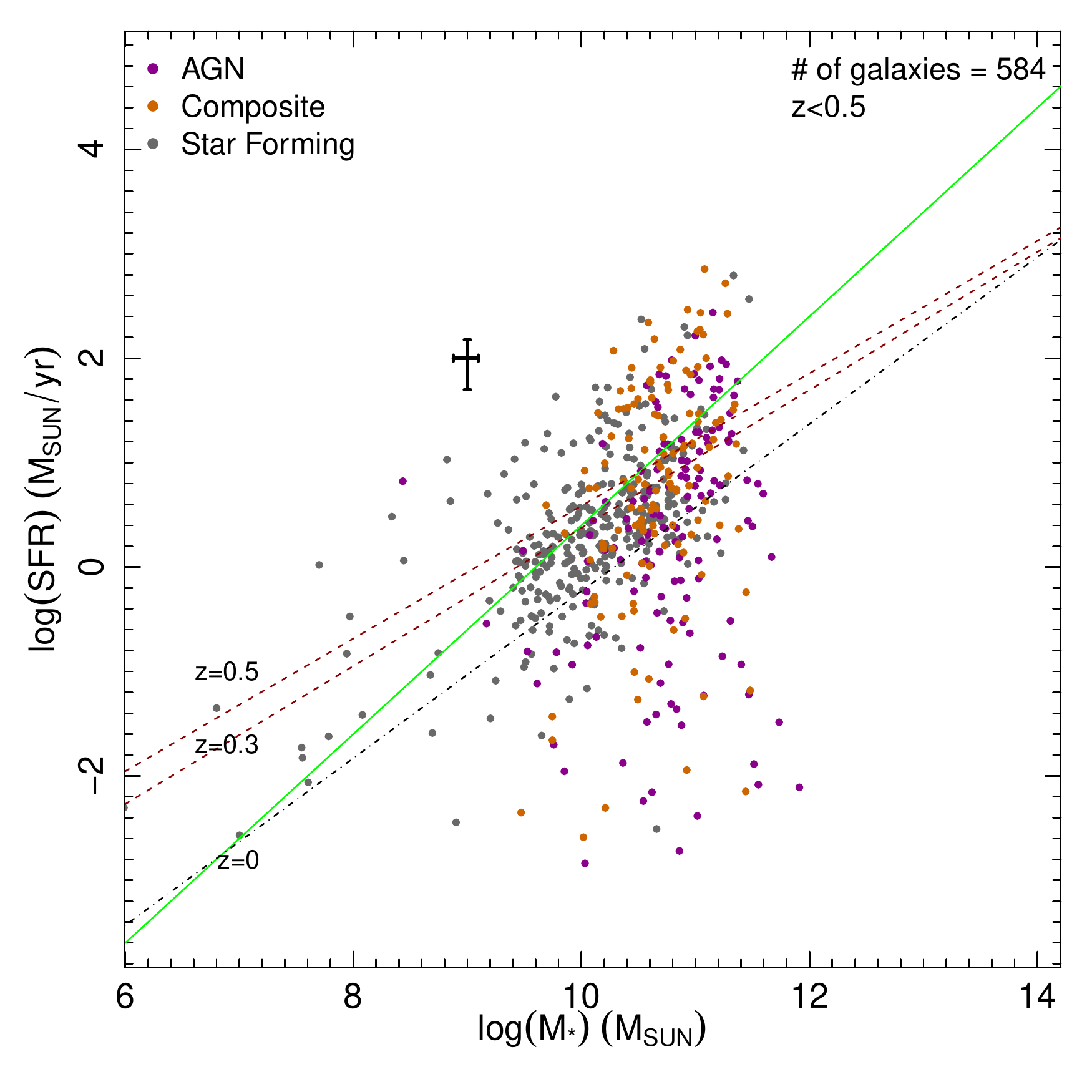}
\caption{The distribution of the 584 galaxies with optical spectral lines on the log($SRF$) - log($M_\star$) plane. 
The colour coding is based on the galaxy classification from Fig. \ref{fig:bpt-agnfr}.}
\label{fig:ms-agnfr}
\end{figure}

\subsection{AGN classification of low- and moderate- redshift samples}

Earlier in this Section we showed how the ${\rm frac}_{\rm AGN}$ that is a derived parameter highlights the AGN population of galaxies in our sample. 
In this section we will rely on optical spectroscopical measurements in order to explore further the presence of AGNs.
For this purpose we employ a version of BPT diagram (\citealt{tex:BP81}) as presented in \citet{tex:KG06}.
In Fig.~\ref{fig:bpt-agnfr} we plot the distribution of 584 galaxies with z$<0.5$ for which we can retrieve line fluxes from the SDSS spectroscopic catalogue\footnote{galSpecLine-dr8 catalogue,  http://www.sdss.org/dr12/spectro/spectro \_access/}.
The BPT diagram separates the galaxies in three groups, star-forming galaxies, AGNs and composite galaxies that are characterised by high star-formation and at the same time host an AGN.

In the top panel of Figure~\ref{fig:bpt-agnfr} we colour code the points based on the ${\rm frac}_{\rm AGN}$ parameter. 
The solid line in this figure corresponds to the theoretical upper limit for pure starburst galaxies. 
Galaxies found at the right part of the solid line have a substantial AGN contribution to the line fluxes.
The dotted line is defined in \citet{tex:KH03c} and is a lower limit for separating AGN galaxies from star forming. 
The $H \alpha$ flux of the galaxies that reside between the two lines might originate, up to 40\%, from an AGN. 
Finally, galaxies below the dotted line are not AGN even by the most conservative AGN criterion defined by the optical emission lines. 
We find that $\sim$70\% of the galaxies that have ${\rm frac}_{\rm AGN}>0.3$ are lying on the AGN and composite region on the BPT diagram.
However there are nine galaxies with ${\rm frac}_{\rm AGN}>0.5$ that are found below the dashed line.
By individually inspecting these galaxies we find that eight out of nine have an AGN classification according to NED type. 
These nine cases are the following: PG 1121+422 (QSO),  SDSS J142748.28+050222.0 (QSO), [HB89] 1552+085 (QSO), 2MASS J09184860+2117170 (Seyfert), 
SDSS J144507.29+593649.8 (Sy1), SDSS J143705.58+611522.5 (no available classification), SDSS J092600.40+442736.1 (AGN), SDSS J171207.44+584754.4 (QSO) and SDSS J231812.99-004126.0 (AGN). 
Additionally, their IRS spectra  do not show any strong PAH features that supports further the existence of AGN emission.

As various studies have already shown (e.g. \citealt{tex:SC06}) the classification based on the BPT diagram requires multiple emission lines in order to be robust. 
The separation between galaxies appears to be not clear for a large population of galaxies which host both star-formation and an AGN and should be treated with caution, see also \citet{tex:BM16}.
On the other hand the advantage of classifying the AGN presence with CIGALE is that one combines information from the UV, optical and IR fluxes and is not limited to the optical emission lines that can be attenuated due to dust. 
This implies that it is possible for a galaxy to host a strong AGN noticeable with SED modelling but not detectable with optical spectra diagnostics.
Another advantage of using the SED modelling AGN classification for this sample is that it provides a classification for the full sample and not only those with optical spectroscopy.

In the bottom panel of Figure~\ref{fig:bpt-agnfr} we colour code the points based on the AGN luminosity as measured by CIGALE.
We notice that there is a gradual increase of the AGN luminosity of the galaxies as we move from the star-forming region to the AGN region and that, on average, galaxies with higher AGN luminosities have AGN that contributes more to the total IR luminosity. 
Examining both BPT diagrams it becomes evident that galaxies in the optically defined starburst region with weak or non existent AGN contribution to their IR luminosity as quantified by the CIGALE fit (${\rm frac}_{\rm AGN}<0.3$ ), 
may have an estimated AGN luminosity  that is moderate, mostly well below $10^{38}$ ergs s$^{-1}$. 
However, strong AGN both based on their optical lines, as well as in the IR (${\rm frac}_{\rm AGN}>0.3$ ), are more luminous typically exceeding $10^{38}$ ergs s$^{-1}$.

There is a large fraction of composite and AGN galaxies that are found to have ${\rm frac}_{\rm AGN}<0.3$.
This result can be the effect of two acts. In the first case these galaxies are classified as composite and AGN with optical spectral lines but CIGALE finds that the AGN luminosity contribution is very low compared to the other components. 
Considering that the SDSS fibers are 3 arcsec in diameter which is much smaller than the typical broad-band photometry apertures used to do the SED fitting,
we can conclude that CIGALE estimates are more realistic as they sample the full physical size of the  extended star-forming circumnuclear regions compared to the SDSS nuclear spectra. 
It is important to remind the reader that the BPT diagram does not tell us something about the AGN contribution to the IR or bolometric luminosity that CIGALE is measuring but only shows that there are high energy optical photons not consistent with star-forming obtained by SDSS.

In the second case, these galaxies do host an AGN but it is not possible to retrieve with SED modelling.
When CIGALE fits AGN emission it is not possible to disentangle between different AGN models, except for very particular cases such as a Type 1 with a strong power law or a deeply enshrouded Type 2. 
Thus the code is looking for a deviation from a ``normal'' galaxy SED rather than a shape of the AGN emission.
Tests showed that there are very few cases that CIGALE has misclassified a galaxy that hosts an AGN to a galaxy with ${\rm frac}_{\rm AGN}<0.3$.

Finally, in Figure \ref{fig:ms-agnfr} we present the distribution of the 584 galaxies with optical spectral lines on the log($SRF$) - log($M_\star$) plane colour-coded based on their classification as defined by the three regions of Fig.~\ref{fig:bpt-agnfr}. 
We find that almost all the galaxies in this sample that have stellar mass less than $10^{10}$ M$_{\odot}$ are classified as star forming. 
We notice that the composite and AGN galaxies do not lie on a particular area of the MS.

\section{Conclusions}  
\label{sec:conclusions}

We have analysed the broadband SEDs of 1,146 galaxies with available mid-IR spectra from Spitzer/IRS and broadband photometry from the UV to 22$\mu$m. 
We have collected photometric measurements from GALEX, SDSS, 2MASS/UKIRT and WISE and fitted the photometric SED with the code CIGALE.

The CASSIS galaxies span a wide redshift range from 5Mpc up to z$\sim$2.5. 
Based on the CIGALE measured parameters it is shown that the local sample (z$<$0.1) consists of 584 galaxies that have a wide range of SFRs and stellar masses, from massive passive galaxies to dwarf galaxies with SFRs above the MS. 
The mid-redshift sample (0.1$ \leq $z$<$0.5) consists of 360 galaxies and the high-redshift  sample (z$\geq $0.5) of 190 galaxies, both samples are dominated by massive, star forming galaxies that are placed above the MS corresponding at each redshift bin. 
Employing additionally the CIGALE parameter ${\rm frac}_{\rm AGN}$ we find that the low redshift galaxies are mainly star forming and only 10\% and 2\% are hosting a moderate and a strong AGN emission respectively. 
The percentage of the composite and AGN galaxies for the mid-redshift and high-redshift increases from 18\% to 29\% and 13\% to 24\%.

The star-formation properties of all the galaxies at z$<$0.5 and available $NUV-r$ restframe colours,  915 in total, are further explored based on the $NUV-r$ colour - absolute $r-$band magnitude diagram. 
With the use of this diagram we confirm that the vast majority (97\%) of the galaxies in this sample have experienced a recent star-formation event in agreement with the  high sSFR as measured by CIGALE.

For a fraction of 256 galaxies with z$<$0.3 with available single S\'ersic index $n_{\rm g}$ measurements the galaxies are divided based on their structure into early-, "red \& low $n$"-, late- and "blue \& high $n$"-type galaxies.
CASSIS galaxies display a wide range of structures and when are placed on the $\log (\rm SFR) - \log (M_{\star})$ plane show a gradual distinction from early-type galaxies to "blue \& high $n$" galaxies that indicates a connection of their structure with the sSFR.
More specific, early-type galaxies are located below the main star forming sequence and "blue \& high $n$" galaxies are found above the MS.
On the contrary, "red \& low $n$" galaxies together with late-type galaxies are settled mainly on the MS with the former to have on average lower SFR and higher stellar mass.

A subsample of 586 galaxies that optical spectral line measurements can be acquired is delineated into AGN, composite and star-forming galaxies based on the BPT diagram.
It is found that the optical spectral line classification is not always in agreement with the CIGALE model parameter ${\rm frac}_{\rm AGN}$ but there is a correlation between the CIGALE AGN luminosity and the optical spectral line classification, 
where AGN luminosity is gradually increasing as we move from star-forming, to composite and AGN galaxies.
We speculate that this mismatch as a result first of the different region for collecting light used by each method, i.e. central 3 arcsec in case of SDSS spectra versus the total light in the case of broadband photometry.
Secondly, it is due to different wavelength range studied by each methodology, SDSS lines investigate only optical signs of AGN or star forming activity while with SED modelling uses the information from UV, optical and IR measurements.
Placing these galaxies on the MS and using the classification acquired from the BPT diagram we see that CASSIS AGN galaxies do not occupy a specific region on the diagram and can be found above, on and below the MS revealing that exhibit various sSFRs.

Finally, this study provides a catalogue with all the CIGALE measured physical parameters along with a structure classification and a star forming - AGN activity classification.
The two classifications are derived from the colour - $n_{\rm g}$ and BPT diagrams respectively. 
The SED derived physical parameters contained in the catalogue are the stellar mass ($M_\star$), the instantaneous star-formation rate (SFR), the stellar age, the E(B-V) attenuation of the young stellar population, the AGN luminosity fraction (${\rm frac}_{\rm AGN}$),  the AGN luminosity (L$_{\rm AGN}$), the dust luminosity (L$_{\rm dust}$) and the dust attenuation in the $FUV$ (A$_{\rm FUV}$). 
It should be stressed that the availability of Spitzer/IRS nuclear spectra for all the galaxies in our sample provides a unique advantage compared to other studies. 
In a subsequent paper we will explore how the nuclear properties of the sample, as derived by spectral features in the rest-frame 5-37 $\mu$m range, which are not (or marginally) affected by obscuration, 
compare with integrated galaxy properties obtained by modelling their global SED.

\section*{Acknowledgments}
We thank Manolis Rovilos for his support in collecting part of the sample.
We also thank the referee, Dr. M. Boquien, whose constructive comments helped improve the manuscript.
This work was performed in the framework of PROTEAS project within GSRT's KRIPIS action, 
funded by Greece and the European Regional Development Fund of the European Union under the O.P. Competitiveness and Entrepreneurship, NSRF 2007-2013 and the Regional Operational Program of Attica.

LC acknowledges the financial support from the EU Seventh Framework Programme (FP7/2007-2013) under grant agreement n. 312725 and the {\sc  thales} project 383549 that  is jointly funded by the European Union and the Greek Government in the framework of the program ``Education and lifelong learning''. 

Funding for SDSS-III has been provided by the Alfred P. Sloan Foundation, the Participating Institutions, the National Science Foundation, and the U.S. Department of Energy Office of Science. 
The SDSS-III web site is http://www.sdss3.org/. SDSS-III is managed by the Astrophysical Research Consortium for the Participating Institutions of the SDSS-III Collaboration including the University of Arizona, the Brazilian Participation Group, Brookhaven National Laboratory, Carnegie Mellon University, University of Florida, the French Participation Group, the German Participation Group, Harvard University, the Instituto de Astrofisica de Canarias, the Michigan State/Notre Dame/JINA Participation Group, Johns Hopkins University, Lawrence Berkeley National Laboratory, Max Planck Institute for Astrophysics, Max Planck Institute for Extraterrestrial Physics, New Mexico State University, New York University, Ohio State University, Pennsylvania State University, University of Portsmouth, Princeton University, the Spanish Participation Group, University of Tokyo, University of Utah, Vanderbilt University, University of Virginia, University of Washington, and Yale University.

This publication makes use of data products from the Two Micron All Sky Survey, which is a joint project of the University of Massachusetts and the Infrared Processing and Analysis Center/California Institute of Technology, funded by the National Aeronautics and Space Administration and the National Science Foundation.

This publication makes use of data products from the Wide-field Infrared Survey Explorer, which is a joint project of the University of California, Los Angeles, and the Jet Propulsion Laboratory/California Institute of Technology, funded by the National Aeronautics and Space Administration.

\bibliography{cassis_cigale}

\onecolumn

\setlength{\textwidth}{6.8in}
\LTcapwidth=\textwidth
\begin{landscape}

\footnotesize 
\begin{center}
\begin{longtable}{l@{\hskip 2mm}c@{\hskip 2mm}c@{\hskip 2mm}l@{\hskip 2mm}l@{\hskip 2mm}l@{\hskip 2mm}l@{\hskip 2mm}l@{\hskip 2mm}l@{\hskip 2mm}l@{\hskip 2mm}l@{\hskip 2mm}l@{\hskip 2mm}c@{\hskip 2mm}c@{\hskip 2mm}l@{\hskip 2mm}c@{\hskip 2mm}c@{\hskip 2mm}c@{\hskip 2mm}}
\caption{Physical parameters of all the Spitzer/IRS galaxies resulting from CIGALE SED fitting together with two galaxy-type classifications from Figures \ref{fig:colour-n} and \ref{fig:bpt-agnfr}. 
The full table is available in a machine-readable form in the online journal and at http://members.noa.gr/mvika/research.php.}\\
\label{table:finalpar}\\

\hline  \\[-2.0ex]
\multicolumn{1}{c}{AORkey } & 
\multicolumn{1}{c}{RA } &
\multicolumn{1}{c}{DEC} &
\multicolumn{1}{c}{z} & 
\multicolumn{1}{c}{Name} & 
\multicolumn{1}{c}{log(M$_\star$)} & 
\multicolumn{1}{c}{SFR} & 
\multicolumn{1}{c}{log(sSFR)} & 
\multicolumn{1}{c}{age} &
\multicolumn{1}{c}{E(B-V)}  &
\multicolumn{1}{c}{${\rm frac}_{\rm AGN}$} & 
\multicolumn{1}{c}{ L$_{\rm AGN}$} &
\multicolumn{1}{c}{L$_{\rm dust}$} & 
\multicolumn{1}{c}{A$_{\rm FUV}$} &
\multicolumn{1}{c}{$\chi^2$} &
\multicolumn{1}{c}{A} &
\multicolumn{1}{c}{B}  &
\multicolumn{1}{c}{f}  \\[0.5ex]

\multicolumn{1}{l}{    } & 
\multicolumn{1}{c}{    } & 
\multicolumn{1}{c}{    } & 
\multicolumn{1}{l}{    } & 
\multicolumn{1}{l}{    } & 
\multicolumn{1}{c}{M$_{\odot}$} & 
\multicolumn{1}{c}{M$_{\odot}$/yr} &
\multicolumn{1}{c}{yr$^{-1}$} &  
\multicolumn{1}{c}{Myr} & 
\multicolumn{1}{c}{    } & 
\multicolumn{1}{c}{    } & 
\multicolumn{1}{c}{  W  } & 
\multicolumn{1}{c}{  W  } & 
\multicolumn{1}{c}{  mag } & 
\multicolumn{1}{c}{    } & 
\multicolumn{1}{c}{    } & 
\multicolumn{1}{c}{     }  \\[0.5ex]

\hline \hline
\endhead

\hline
\endfoot

\hline
\endlastfoot\\
15077888  & 0.7  & 0.75  & 0.09  & 2MASX\_J00024910+0045055  & 11.3, 10.6  & 0.3, 0.4  & -11.8, -11.7  & 7463, 2264  & 0.25, 0.08  & 0.19, 0.17  & 2e+36, 1.6e+36  & 8e+36, 3e+36  & 1.6  & 0.9  & 4  & 1  & 0 \\
14188288  & 1.5  & 16.16  & 0.45  & PG\_0003+158  & 11.7, 10.9  & 764, 200.8  & -8.8, -9.4  & 1222, 945  & 0.01, 0.02  & 0.78, 0.12  & 6e+38, 9.7e+37  & 2e+38, 2e+38  & 0.1  & 2.7  & 0  & 0  & 0 \\
15079680  & 10.72  & 15.55  & 0.12  & 2MASX\_J00425255+1532465  & 11.2, 10.6  & 21.8, 12.7  & -9.9, -10.1  & 4325, 2607  & 0.47, 0.13  & 0.27, 0.15  & 2.5e+37, 8.8e+36  & 7e+37, 3e+37  & 3  & 1.3  & 0  & 1  & 0 \\
14029312  & 11.22  & -26.89  & 0.12  & 2MASS\_J00445167-2653191  & 10.4, 9.9  & 5.1, 2.5  & -9.7, -9.9  & 5220, 2991  & 0.49, 0.13  & 0.06, 0.04  & 3e+36, 1.4e+36  & 2e+37, 3e+36  & 2.2  & 0.3  & 0  & 0  & 0 \\
10951168  & 11.45  & 4.17  & 0.39  & PG\_0043+039  & 11.6, 10.7  & 143.6, 156.2  & -9.4, -9.4  & 1062, 321  & 0.1, 0.01  & 0.52, 0.08  & 4.4e+38, 6e+37  & 3e+38, 1e+38  & 0.5  & 1.8  & 0  & 0  & 0 \\
15075584  & 11.83  & 14.7  & 0.04  & UGC\_00488  & 11.2, 10.6  & 4.2, 2.6  & -10.5, -10.7  & 7332, 3328  & 0.29, 0.1  & 0.31, 0.14  & 8.6e+36, 3e+36  & 2e+37, 8e+36  & 1.9  & 0.4  & 0  & 3  & 0 \\
24412672  & 110.47  & 71.34  & 0.3  & [HB89]\_0716+714  & 12.9, 12.3  & 1771.7, 852.9  & -9.6, -9.9  & 5206, 2389  & 0.2, 0.06  & 0.41, 0.1  & 3e+39, 3.6e+38  & 3e+39, 1e+39  & 1.3  & 0.3  & 0  & 0  & 0 \\
15071232  & 119.29  & 42.6  & 0.07  & 2MASX\_J07570930+4236163  & 10.6, 10  & 1, 0.8  & -10.6, -10.7  & 7757, 2949  & 0.27, 0.06  & 0.08, 0.05  & 1.2e+36, 8.3e+35  & 5e+36, 1e+36  & 1.7  & 0.8  & 0  & 2  & 0 \\
14022656  & 119.5  & 39.34  & 0.1  & FBQS\_J075800.0+392029  & 11.4, 10.9  & 67.9, 34.5  & -9.6, -9.8  & 4940, 2618  & 0.22, 0.05  & 0.41, 0.08  & 1.2e+38, 1.2e+37  & 1e+38, 5e+37  & 1.4  & 1.1  & 0  & 0  & 1 \\
14807296  & 119.62  & 37.79  & 0.04  & NGC\_2484  & 11.9, 10.7  & 0, 0  & -14, -13.7  & 8716, 842  & 0.01, 0  & 0.49, 0.14  & 1.3e+36, 5.1e+35  & 1e+36, 7e+34  & 0  & 0.1  & 0  & 1  & 0 \\
14189056  & 12.98  & 17.43  & 0.06  & MRK\_1148  & 10, 9.4  & 25.1, 5  & -8.6, -9.1  & 1978, 1646  & 0.01, 0  & 0.6, 0.03  & 6.5e+36, 7.9e+35  & 3e+36, 4e+35  & 0.1  & 2.1  & 0  & 0  & 0 \\
11297792  & 121.4  & 24.16  & 0.06  & 3C\_192  & 11.2, 10.5  & 0.1, 0.1  & -12.1, -12.1  & 7884, 1792  & 0.1, 0.02  & 0.11, 0.14  & 6.2e+35, 4.1e+35  & 2e+36, 5e+35  & 0.7  & 0.5  & 4  & 1  & 1 \\
15075328  & 127.3  & 50.11  & 0.04  & SDSS\_J082912.67+500652.3  & 9.6, 9.1  & 1.1, 0.5  & -9.5, -9.8  & 5356, 2920  & 0.23, 0.08  & 0.23, 0.13  & 8.5e+35, 3.9e+35  & 2e+36, 8e+35  & 1.7  & 0.8  & 2  & 3  & 0 \\
15070464  & 129.52  & 54.11  & 0.03  & SDSS\_J083803.67+540642.1  & 9.5, 8.9  & 0.1, 0  & -10.5, -10.8  & 8391, 3296  & 0.11, 0.03  & 0.12, 0.09  & 5.7e+34, 4.6e+34  & 2e+35, 5e+34  & 0.8  & 1.3  & 0  & 3  & 0 \\
10869760  & 129.55  & 24.9  & 0.03  & NGC\_2622  & 11.2, 10.5  & 1.8, 1.4  & -10.9, -11  & 7121, 3341  & 0.44, 0.09  & 0.07, 0.08  & 1.3e+36, 1.4e+36  & 2e+37, 5e+36  & 2.5  & 0.3  & 0  & 1  & 0 \\
12239872  & 130.2  & 13.21  & 0.68  & 3C\_207  & 11.7, 11.1  & 190.4, 147.4  & -9.4, -9.5  & 1971, 1231  & 0.27, 0.1  & 0.49, 0.18  & 5.3e+38, 8.5e+37  & 5e+38, 3e+38  & 1.2  & 3.9  & 0  & 0  & 0 \\
15075072  & 130.41  & 54.92  & 0.04  & 2MASX\_J08413787+5455069  & 10.8, 10.3  & 0.1, 0.2  & -11.7, -11.4  & 5338, 2834  & 0.26, 0.14  & 0.6, 0.17  & 8.1e+36, 1.1e+36  & 5e+36, 3e+36  & 0.7  & 0.3  & 0  & 1  & 0 \\
10482944  & 130.79  & 29.73  & 0.4  & 4C\_+29.31  & 11.2, 10.8  & 96.5, 69.7  & -9.2, -9.3  & 5744, 2308  & 0.45, 0.17  & 0.46, 0.17  & 1.7e+38, 2e+37  & 2e+38, 1e+38  & 2  & 0.9  & 0  & 0  & 0 \\
15074560  & 133.98  & 0.85  & 0.05  & 2MASX\_J08555426+0051110  & 10.6, 9.9  & 2, 1  & -10.3, -10.5  & 9150, 2951  & 0.18, 0.06  & 0.47, 0.09  & 5.4e+36, 5.8e+35  & 5e+36, 2e+36  & 0.8  & 0.6  & 0  & 1  & 1 \\
15076864  & 136.91  & 56.73  & 0.1  & SDSS\_J090738.71+564358.2  & 10.5, 9.9  & 1.8, 1  & -10.3, -10.5  & 5542, 2921  & 0.28, 0.08  & 0.21, 0.13  & 2.1e+36, 1.2e+36  & 7e+36, 2e+36  & 2  & 0.9  & 0  & 1  & 0 \\
15073792  & 137.76  & 57.21  & 0.05  & 2MASX\_J09110135+5712472  & 10.6, 10  & 0, 0.1  & -12.1, -11.8  & 7009, 2795  & 0.28, 0.1  & 0.28, 0.17  & 8.3e+35, 4.6e+35  & 2e+36, 9e+35  & 1.3  & 0.2  & 0  & 1  & 0 \\
15082496  & 139.98  & 55.36  & 0.12  & MRK\_0106  & 10.4, 10.3  & 65.7, 27.3  & -8.6, -8.7  & 5019, 3424  & 0.11, 0.04  & 0.42, 0.1  & 1.6e+38, 2.5e+37  & 6e+37, 2e+37  & 1  & 0.8  & 0  & 3  & 0 \\
13628928  & 14.09  & 0.54  & 0.48  & SDSS\_J0056+0032  & 11, 10.5  & 146.2, 112.8  & -8.8, -8.9  & 5168, 2140  & 0.64, 0.15  & 0.35, 0.27  & 1.2e+38, 8.3e+37  & 2e+38, 1e+38  & 3.8  & 1.1  & 0  & 0  & 0 \\
14805504  & 14.4  & -1.39  & 0.05  & UGC\_00595  & 11.7, 10.5  & 0, 0  & -13.7, -13.7  & 7124, 566  & 0.01, 0  & 0.43, 0.25  & 9.9e+35, 9.9e+35  & 7e+35, 5e+34  & 0.1  & 1.1  & 0  & 0  & 0 \\
13628416  & 140.06  & 45.53  & 0.4  & SDSS\_J092014.11+453157.2  & 11.1, 10.8  & 95.6, 72.6  & -9.1, -9.2  & 4615, 2528  & 0.46, 0.14  & 0.41, 0.16  & 1.4e+38, 1.8e+37  & 2e+38, 1e+38  & 3.8  & 0.7  & 0  & 0  & 0 \\
11298048  & 140.29  & 45.65  & 0.17  & 3C\_219  & 11.6, 11  & 5, 4.6  & -10.9, -10.9  & 6164, 2203  & 0.31, 0.15  & 0.47, 0.2  & 3.2e+37, 1.3e+37  & 3e+37, 2e+37  & 2.5  & 0.9  & 0  & 1  & 0 \\
14203392  & 141.51  & 12.73  & 0.03  & UGC\_05025  & 10.8, 10.3  & 14.2, 6.4  & -9.6, -9.9  & 7427, 3346  & 0.45, 0.2  & 0.16, 0.1  & 8.9e+36, 3.2e+36  & 3e+37, 1e+37  & 2.6  & 0.4  & 0  & 3  & 0 \\
15072512  & 143.31  & 1.98  & 0.03  & SDSS\_J093313.94+015858.7  & 8.9, 8.1  & 0, 0  & -11.3, -10.9  & 3536, 1369  & 0.01, 0.01  & 0.27, 0.19  & 1e+34, 1.2e+34  & 1e+34, 9e+33  & 0  & 0.4  & 1  & 3  & 0 \\
15082240  & 145.79  & 60.77  & 0.07  & SDSS\_J094310.11+604559.1  & 9.7, 8.9  & 0.2, 0.5  & -10.3, -10  & 1645, 1005  & 0.34, 0.08  & 0.35, 0.11  & 1.8e+36, 2.3e+35  & 3e+36, 1e+36  & 1.8  & 0.7  & 0  & 3  & 0 \\
14190592  & 147.7  & 39.45  & 0.21  & PG\_0947+396  & 11.1, 11  & 168.6, 112.8  & -8.8, -8.8  & 5935, 3525  & 0.29, 0.34  & 0.69, 0.15  & 3.8e+38, 2e+38  & 7e+37, 6e+37  & 0.1  & 1.5  & 0  & 2  & 0 \\
\end{longtable}   
\end{center}
\normalsize
\tablefoot{Column 1:  Spitzer/IRS identification key, 
Columns 2 \& 3: coordinates of the Spitzer/IRS source, 
Column 4: redshift from NED, 
Column 5: galaxy name from NED, 
Column 6: stellar mass as measured by CIGALE, 
Column 7: star-formation rate as measured by CIGALE, 
Column 8: specific star-formation rate,
Column 9: stellar age as measured by CIGALE, 
Column 10: E(B-V) attenuation of young stellar populations as measured by CIGALE, 
Column 11: the contribution of an AGN to the total IR luminosity as measured by CIGALE, 
Column 12: AGN luminosity as measured by CIGALE, 
Column 13: dust luminosity as measured by CIGALE, 
Column 14: the attenuation in the $FUV$ as measured by CIGALE, 
Column 15: minimised chi square of the fit, 
Column 16: classification based on Fig.\ref{fig:colour-n} where 1 indicates late-, 2 blue \& high- $n$, 3 red \& low- $n$, 4 early-type galaxy and 0 not available classification, 
Column 17: classification based on Fig.\ref{fig:bpt-agnfr} where 1 indicates AGN, 2 composite, 3 star-forming galaxy and 0 not available classification, 
Column 18: flag where 1 indicates saturation flag in SDSS data.
Where two values are provided in a column the second is the uncertainty provided by CIGALE code.
Galaxies with AORkey 28146432, 24189952, 26906368, 15510272, 18619392, 11350016 and 18600448 have a mismatch with IDEOS redshift measurements, see text for more details. 
While, the galaxies with AORkey 14022656, 14194688 and 28975872 show signs of saturation in their optical images that possible disturbs their photometric measurements.
The CIGALE stellar age should be used with caution, see text for more details.
} 
\end{landscape}

\label{lastpage}

\end{document}